# Resolved SED modelling with *JWST* and ALMA: the role of stellar mass surface density in regulating star formation in cosmic noon galaxies

Cheng-Lin Liao (廖政霖),[1★] Leindert A. Boogaard,[1★] Jacqueline A. Hodge,[1★] J. Li,[2,3,4] E. da Cunha,[2,3] P. Sharda,[1] A. Battisti,[2,3,5] C. Lagos,[2,3] M. Kaasinen,[5] I. Smail,[6] M. Aravena,[7,8] F. E. Bauer,[9] R. K. Cochrane,[10] L. Colina,[11] P. Cox,[12] R. Decarli,[13] T. Díaz-Santos,[14] R. Fernández Aranda,[11] N. N. Geesink,[15] H. Inami,[16] B. Magnelli,[17] J. Melinder,[18] M. Neeleman,[19] G. Östlin,[18] G. Popping,[15] D. A. Riechers,[20] I. Shivaei,[11] P. van der Werf,[1] F. Walter[21] and A. Weiss[22]

*Affiliations are listed at the end of the paper*

Accepted 2026 September 4. Received 2026 September 3; in original form 2026 May 1

## ABSTRACT

We present kpc-scale (0.2–0.5 arcsec) physical property maps of 35 main-sequence galaxies at $z \approx 0.5$–3.7, with stellar masses of $\log(M_*/\mathrm{M}_\odot) \sim 9.7$-11.7 and star formation rates of SFR $\sim 1.4$-280 $\mathrm{M}_\odot\,\mathrm{yr}^{-1}$, selected from the ALMA Spectroscopic Survey (ASPECS) in the Hubble Ultra Deep Field. Leveraging the unique *Hubble Space Telescope*, *James Webb Space Telescope* with the Near Infrared Camera (NIRCam) and Mid-Infrared Instrument (MIRI), and ALMA observations, we perform spatially resolved spectral energy distribution (SED) modelling across the ultraviolet to far-infrared regime. We find that incorporating MIRI and/or ALMA data reduces the overestimation of dust luminosity (by up to $\sim 0.8$ dex), while ALMA observations further mitigate the age-dust degeneracy. In the absence of such data, restricting the SED model library based on the observed unresolved colours can partially mitigate these biases. The stellar masses ($M_*$) derived from resolved and unresolved modelling are consistent within $\sim 0.05$ dex, suggesting that mass discrepancies (attributed to outshining) are less significant for cosmic noon main-sequence galaxies when rest-frame near-infrared (e.g. $\sim 1$–3 µm) data are included. After normalization to the same reference, the composite SED of our sample closely resembles that of local starburst galaxies such as M82, suggesting similar dust attenuation and re-emission properties. Finally, we find that the molecular gas fraction and depletion time correlate with the effective stellar mass surface density ($\Sigma_{\mathrm{eff},*} = M_*/2\pi R^2_{\mathrm{eff,M_*}}$) similarly to that observed in local galaxies. These results provide a first qualitative view of how the stellar gravitational potential influences gas regulation and star formation in galaxies beyond the local Universe.

## 1 INTRODUCTION

The cosmic star formation rate (SFR) density peaks ~10 billion years ago, an epoch commonly referred to as 'cosmic noon' ($z \approx 1$–3; P. Madau & M. Dickinson 2014). Compared to the local Universe, galaxies at these redshifts are more gas-rich (e.g. R. Decarli et al. 2019; L. A. Boogaard et al. 2023), with roughly an order of magnitude higher molecular to stellar mass ratios (e.g. L. J. Tacconi et al. 2010; L. J. Tacconi, R. Genzel & A. Sternberg 2020), and exhibit elevated SFRs at fixed stellar mass ($M_*$; J. S. Speagle et al. 2014; K. E. Whitaker et al. 2014; P. Popesso et al. 2023). In addition, the molecular gas depletion time ($t_{\mathrm{depl,mol}}$) appears to evolve with redshift, with higher-redshift galaxies displaying slightly shorter $t_{\mathrm{depl,mol}}$ (e.g. J. A. Hodge & E. da Cunha 2020; L. J. Tacconi et al. 2020; F. Walter et al. 2020). While these trends have been largely established from global measurements, how they are impacted by the internal structure of galaxies is still poorly understood. In particular, the spatial distributions of key physical quantities, such as the molecular gas surface density ($\Sigma_{\mathrm{gas,mol}}$) and stellar mass surface density ($\Sigma_*$), remain largely unconstrained for typical galaxies at cosmic noon.

A common approach to explore star formation across cosmic time has relied on the Kennicutt–Schmidt relation (M. Schmidt 1959; R. C. Kennicutt 1998), which links the SFR directly to the gas abundance. While this empirical relation has been investigated extensively, it only focuses on the contribution from gas, without considering the additional regulation provided by the underlying stellar mass. Within the framework of galactic

★ E-mail: liao@strw.leidenuniv.nl (CLL); boogaard@strw.leidenuniv.nl (LAB); hodge@strw.leidenuniv.nl (JAH)

disc formation models (e.g. M. R. Krumholz et al. 2018; E. C. Ostriker & C.-G. Kim 2022), $\Sigma_*$, though often treated implicitly, contributes to the gravitational potential that determines the gas scale height (e.g. E. C. Ostriker & R. Shetty 2011; J. Forbes, M. Krumholz & A. Burkert 2012). Since the gas scale height is set by the balance between gravity and turbulence, with the latter largely driven by stellar feedback, $\Sigma_*$ plays a central role in regulating star formation on galactic scales. Consistent with this picture, recent studies of local galaxies have shown that a combination of the effective stellar mass surface density ($\Sigma_{\rm eff,*}$) and the effective gas surface density ($\Sigma_{\rm eff,gas}$) correlates more tightly with the SFR surface density than $\Sigma_{\rm eff,gas}$ alone (e.g. Y. Shi et al. 2011; J. Sun et al. 2020, 2023; E. Schinnerer & A. K. Leroy 2024). Here, $\Sigma_{\rm eff,*}$, defined as the stellar mass surface density within the effective radius (typically taken as the half-light radius), serves as a proxy for the stellar gravitational potential on galactic scales. By accounting for this stellar contribution to the gravitational potential, the regulation of gas collapse and star formation can be more accurately described. Despite this, the role of $\Sigma_{\rm eff,*}$ in regulating star formation beyond the local Universe remains poorly explored, primarily due to the lack of spatially resolved stellar mass measurements at high redshift.

With the advent of the *James Webb Space Telescope* (*JWST*), recent studies have begun to explore spatially resolved physical properties of statistically representative galaxy samples at high redshift using resolved spectral energy distribution (SED) modelling (e.g. C. Giménez-Arteaga et al. 2023, 2024; L. Shen et al. 2024; N. E. P. Lines et al. 2025). Most of these works have focused on rest-frame optical wavelengths enabled by the combination of the *Hubble Space Telescope* (*HST*) and *JWST*/NIRCam imaging. Several studies have reported systematic discrepancies between stellar masses derived from spatially unresolved and spatially resolved modelling, a phenomenon previously discussed and identified in earlier works (L. Searle, W. L. W. Sargent & W. G. Bagnuolo 1973; M. Sawicki & H. K. C. Yee 1998; S. Wuyts et al. 2012; R. Sorba & M. Sawicki 2015, 2018). This discrepancy, often referred to as the 'outshining' effect, describes that the integrated light of galaxies is dominated by young, luminous stellar populations, hiding the true contribution of older and more massive stars. Such biases may be mitigated when rest-frame near-infrared (NIR) data are included, as suggested by J. Song et al. (2023).

Only a limited number of spatially resolved SED modelling studies at high redshift have extended the analyses beyond the rest-frame optical. One of the few examples is J. Li et al. (2024), who analysed ultraviolet (UV)-selected ALMA-CRISTAL galaxies at $4 < z < 6$ by incorporating longer-wavelength data from high-resolution ALMA observations. They found that including FIR constraints in spatially resolved modelling is crucial for deriving robust physical parameter maps, as it enables a better characterization of dust emission and helps to break the age–dust degeneracy. While such studies have demonstrated that wavelength coverage is critical in spatially resolved SED modelling of select high-redshift galaxy populations, it remains unclear whether similar conclusions apply to the typical main-sequence galaxy population at cosmic noon. Moreover, systematic investigations into how varying wavelength coverages quantitatively influence spatially resolved SED fitting results remain limited. A systematic test across different wavelength setups is therefore crucial to benchmark the impact of filter choices on derived resolved properties.

In this study, we focus on the galaxies drawn from the ALMA Spectroscopic Survey in the Hubble Ultra Deep Field (ASPECS) Large Program (F. Walter et al. 2016; R. Decarli et al. 2019; J. González-López et al. 2019; R. Decarli et al. 2020), a well-characterized sample of gas-rich star-forming galaxies at cosmic noon that lie on the star formation main sequence (SFMS). The ASPECS sample benefits from deep multiwavelength coverage (see L. A. Boogaard et al. 2019; M. Aravena et al. 2020), now including *JWST* NIRCam and MIRI observations (G. H. Rieke et al. 2024; G. Östlin et al. 2025; D. J. Eisenstein et al. 2026), see L. A. Boogaard et al. (2024) that provide continuous sampling from the rest-frame UV to mid-infrared (MIR), enabling robust constraints on $M_*$ distributions through the inclusion of rest-frame NIR data. In addition, a small number of ASPECS galaxies also have high-resolution ALMA continuum observations at angular resolutions comparable to those of MIRI at F1280W ($\lesssim 0.5''$; e.g. M. Kaasinen et al. 2020). Such FIR data are crucial for directly constraining dust emission and for enabling spatially resolved UV-to-FIR SED modelling. These characteristics make ASPECS a unique and well-suited sample beyond the local Universe that enables a detailed investigation of the internal baryonic structure of typical galaxies at cosmic noon. Leveraging this unique data set, we explore the role of $\Sigma_{\rm eff,*}$ in regulating star formation in gas-rich galaxies during the peak epoch of galaxy growth ($z \sim 2$). Furthermore, by defining distinct wavelength coverage configurations, we systematically quantify how filter choices impact the derived spatially resolved SED modelling results.

This paper is organized as follows. In Section 2, we present the *JWST*, ALMA, and ancillary data of our sample. In Section 3, we present the analysis that we performed to derive the physical properties. We demonstrate our SED modelling results in Section 4 and discuss in Section 5. In Section 6, we summarize some key points in this paper. Throughout this paper, we adopt a flat $\Lambda$CDM cosmological model with $H_0 = 67.7\,{\rm km\,s^{-1}\,Mpc^{-1}}$, $\Omega_{\rm M} = 0.31$, and $\Omega_\Lambda = 0.69$ (Planck Collaboration VI 2020), and the G. Chabrier (2003) initial mass function (IMF).

## 2 OBSERVATION AND DATA

### 2.1 Sample

Our sample is drawn from the ASPECS Large Program, of which a brief overview and relevant results are provided below. More detailed descriptions can be found in M. Aravena et al. (2019), L. A. Boogaard et al. (2019), R. Decarli et al. (2019), J. González-López et al. (2019), M. Aravena et al. (2020), R. Decarli et al. (2020), and J. González-López et al. (2020). The ASPECS Large Program was conducted during ALMA Cycle 4 (PID: 2016.1.00324.L; PI: F. Walter) as a ∼150 hr blind survey in the Hubble Ultra Deep Field (HUDF), using ALMA Bands 3 and 6 (covering 84–115 GHz and 212–272 GHz; R. Decarli et al. 2019). The programme followed the earlier Cycle 2 ASPECS-Pilot survey (PID: 2013.1.00146.S; PI: F. Walter; M. Aravena et al. 2016; F. Walter et al. 2016; R. Decarli et al. 2016a, b), but with a larger surveyed area of $\sim 5\,{\rm arcmin}^2$, aiming to provide an unbiased measurement of the gas and dust properties of galaxies across cosmic time. The Band 3 observations yielded six 3 mm continuum sources and 16 CO line emitters (L. A. Boogaard et al. 2019; J. González-López et al. 2019), while the Band 6 observations revealed 35 1.2 mm continuum sources, 32 of which have NIR counterparts identified in *HST* imaging (M. Aravena et al. 2020; J. González-López et al. 2020).

In this work, we select our targets from the NIR-confirmed ASPECS galaxies with detections in either the 1.2 mm continuum map or CO emission lines. Our final sample thus consists of 35 galaxies, with colour composite cut-outs shown in Fig. 1. All sources have spectroscopic redshifts confirmed through molecular (L. A. Boogaard et al. 2019, 2020; D. A. Riechers et al. 2020) and/or ionized gas emission lines (H. Inami et al. 2017; L. A. Boogaard et al. 2019; R. Bacon et al. 2023; P. A. Oesch et al. 2023; M. B. Bagley et al. 2024; N. Pirzkal et al. 2024), as summarized in table 1 of L. A. Boogaard et al. (2024). The redshifts span $z \sim 0.5$ to $z \sim 3.7$, with the median of $z = 1.8$. The key properties of our sample are listed in Table 1.

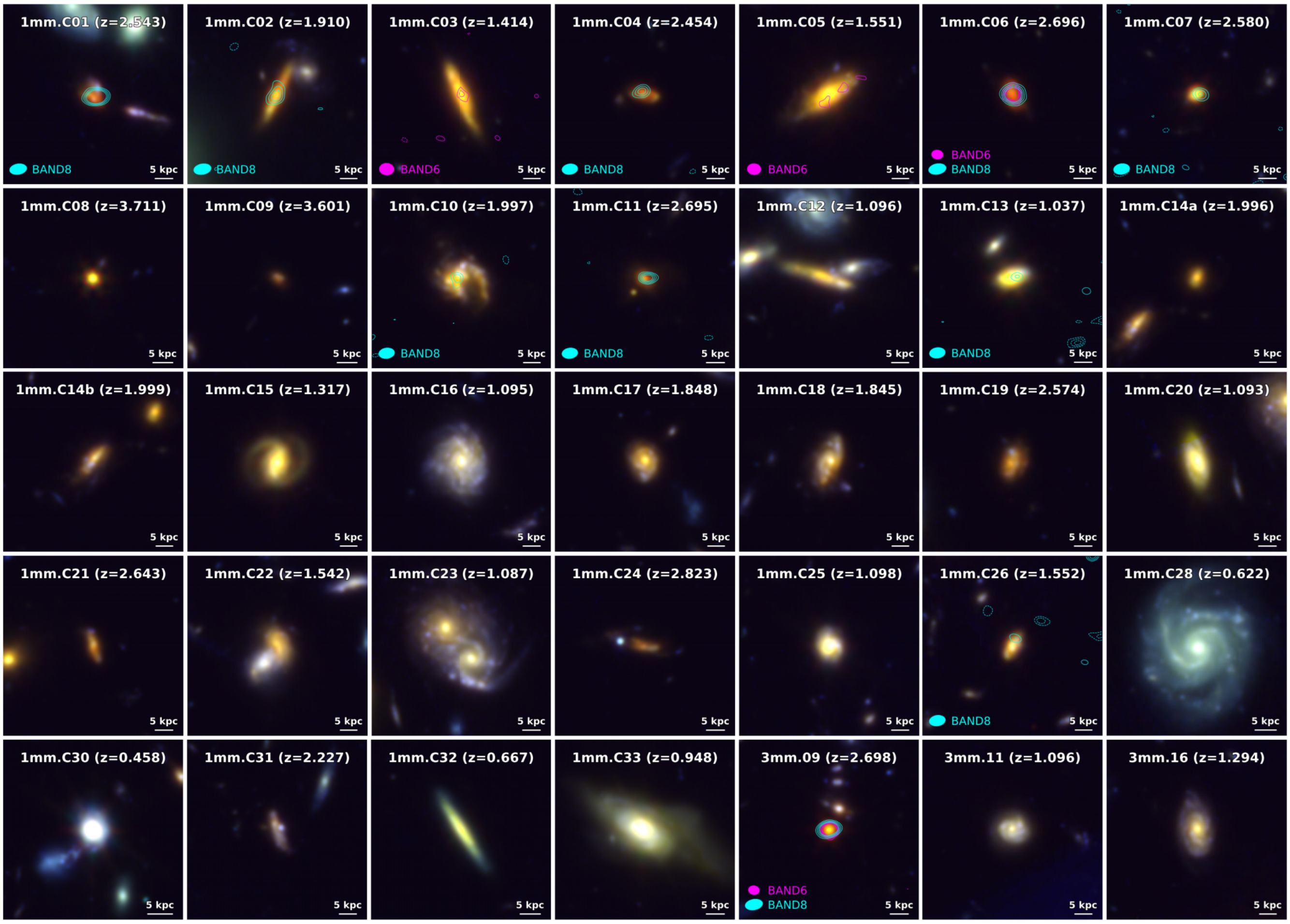

**Figure 1.** Colour-composite cut-outs of all 35 galaxies in our sample, displayed using an asinh stretch. Where available, the high-resolution ALMA Band 8 and 6 continuum emission is overlaid as cyan and magenta contours, respectively. The contours are shown at 3, 4, 5, and 6 $\sigma$ levels, and the synthesized beam sizes are indicated in the lower corners of each panel. The RGB images are constructed from *JWST* NIRCam F444W, NIRCam F277W, and NIRCam F090W for the red, green, and blue channels, respectively. Each cut-out spans 6 arcsec $\times$ 6 arcsec for sources at $z > 1$ and 10 arcsec $\times$ 10 arcsec for those at $z \leq 1$, with a 5 kpc scale bar shown at the bottom right.

Fig. 2 shows their position on the $M_*$–SFR plane overlaid with the SFMS from P. Popesso et al. (2023) at several redshifts, with most of the sources distributed across the SFMS. $M_*$ and SFR are derived from SED modelling with the high-$z$ extension of MAGPHYS (E. da Cunha, S. Charlot & D. Elbaz 2008; E. da Cunha et al. 2015; A. J. Battisti et al. 2020; see more details in Section 3.2). According to L. A. Boogaard et al. (2024), most sources are non-AGN or have only weak AGN contributions, with AGN contributing $< 15$ per cent of the total IR luminosity. Only two sources (1mm.C08 and 3mm.C09) are identified as point-source AGNs and show substantial AGN contributions. While we perform analysis for all sources for completeness, these two point-source AGNs are excluded from all subsequent physical interpretations. For the remaining sources with weak AGN signatures, their AGN contribution to the MIR continuum is sub-dominant and is not expected to significantly bias the derived physical properties. A detailed quantitative decomposition of the AGN contributions is beyond the scope of this work and is deferred to future studies.

### 2.2 *HST* and *JWST* data

To model the SEDs, we use optical data from *HST*, as well as NIR and MIR data from the *JWST*, covering wavelengths from $\sim 0.4$ to $\sim 26\,\mu$m in the observed frame. These data sets were retrieved from the Dawn *JWST* Archive (DJA), an initiative of the Cosmic Dawn Center (DAWN). A summary of the imaging data properties is presented in Table 2.

We include the v7.0 GOODS-South mosaic release for the *HST* filters. The data reduction was performed using the `grizli` pipeline (G. Brammer 2023), which combined *HST* observations from various programs into a mosaic with a pixel scale of 40 milliarcseconds (mas). In this study, we use the DJA-reduced mosaics for all wide-band ACS/WFC filters (F435W, F606W, F775W, F814W, F850LP) in our SED modelling.

**Table 1.** Source properties.

| ID | RA | Decl. | $z_{\rm spec}$ | Resolved in | | High-res. ALMA |
|---|---|---|---|---|---|---|
| | (J2000) | (J2000) | | F444W | F1280W | |
| (1) | (2) | (3) | (4) | (5) | (6) | (7) |
| 1mm.C01 | 03:32:38.55 | −27:46:34.48 | 2.543 | T | F | Band 8 |
| 1mm.C02 | 03:32:36.97 | −27:47:27.20 | 1.910 | T | T | Band 8 |
| 1mm.C03 | 03:32:34.44 | −27:46:59.79 | 1.414 | T | T | Band 6 |
| 1mm.C04 | 03:32:41.01 | −27:46:31.65 | 2.454 | T | F | Band 8 |
| 1mm.C05 | 03:32:39.74 | −27:46:11.57 | 1.551 | T | F | Band 6 |
| 1mm.C06 | 03:32:43.53 | −27:46:39.25 | 2.696 | T | F | Bands 6 and 8 |
| 1mm.C07 | 03:32:35.08 | −27:46:47.79 | 2.580 | F | F | Band 8 |
| 1mm.C08 | 03:32:38.03 | −27:46:26.58 | 3.711 | F | F | |
| 1mm.C09 | 03:32:35.56 | −27:47:04.14 | 3.601 | T | F | |
| 1mm.C10 | 03:32:40.06 | −27:47:55.74 | 1.997 | T | T | Band 8 |
| 1mm.C11 | 03:32:43.33 | −27:46:46.97 | 2.695 | T | F | Band 8 |
| 1mm.C12 | 03:32:36.47 | −27:46:31.98 | 1.096 | T | T | |
| 1mm.C13 | 03:32:42.99 | −27:46:50.22 | 1.037 | T | F | Band 8 |
| 1mm.C14a | 03:32:41.69 | −27:46:55.68 | 1.996 | T | F | |
| 1mm.C14b | 03:32:41.84 | −27:46:57.31 | 1.999 | T | T | |
| 1mm.C15 | 03:32:42.38 | −27:47:07.86 | 1.317 | T | F | |
| 1mm.C16 | 03:32:39.89 | −27:47:15.30 | 1.095 | T | T | |
| 1mm.C17 | 03:32:38.81 | −27:47:14.96 | 1.848 | T | F | |
| 1mm.C18 | 03:32:37.36 | −27:46:45.72 | 1.845 | T | T | |
| 1mm.C19 | 03:32:36.18 | −27:46:27.95 | 2.574 | T | F | |
| 1mm.C20 | 03:32:35.78 | −27:46:27.80 | 1.093 | T | T | |
| 1mm.C21 | 03:32:35.99 | −27:47:25.96 | 2.643 | T | F | |
| 1mm.C22 | 03:32:37.60 | −27:47:43.99 | 1.542 | T | F | |
| 1mm.C23 | 03:32:35.52 | −27:46:26.39 | 1.087[a] | T | T | |
| 1mm.C24 | 03:32:38.74 | −27:48:10.55 | 2.823 | T | T | |
| 1mm.C25 | 03:32:34.86 | −27:46:40.75 | 1.098 | T | F | |
| 1mm.C26 | 03:32:34.69 | −27:46:44.78 | 1.552 | T | F | Band 8 |
| 1mm.C28 | 03:32:40.79 | −27:46:15.97 | 0.622 | T | T | |
| 1mm.C30 | 03:32:38.78 | −27:47:32.39 | 0.458 | F | F | |
| 1mm.C31 | 03:32:37.08 | −27:46:17.46 | 2.227 | T | F | |
| 1mm.C32 | 03:32:37.74 | −27:47:07.19 | 0.667 | T | T | |
| 1mm.C33 | 03:32:38.49 | −27:47:02.69 | 0.948 | T | T | |
| 3mm.09 | 03:32:44.03 | −27:46:35.96 | 2.698 | F | F | Bands 6 and 8 |
| 3mm.11 | 03:32:39.82 | −27:46:53.80 | 1.096 | T | F | |
| 3mm.16 | 03:32:39.92 | −27:46:07.20 | 1.294 | T | T | |

*Notes.*
(1) ALMA 1 mm source ID (1mm.C∗: M. Aravena et al. 2020; J. González-López et al. 2020; 3mm.∗: L. A. Boogaard et al. 2019; J. González-López et al. 2019).
(2) Right Ascension (all values from the DJA catalogue, except for 1mm.C23, which is from JADES; M. J. Rieke et al. 2023).
(3) Declination (all values from the DJA catalogue, except for 1mm.C23, which is from JADES; M. J. Rieke et al. 2023).
(4) Spectroscopic redshift (L. A. Boogaard et al. 2019, 2024).
(5) Robustly resolved in F444W or not ($R_{\rm F444W}/R_{\rm PSF,F444W} \geq 2$): T (True) and F (False).
(6) Robustly resolved in F1280W or not ($R_{\rm F1280W}/R_{\rm PSF,F1280W} \geq 2$): T (True) and F (False).
(7) High-resolution ALMA band (Section 2.3).
[a] 1mm.C23 has a CO-based redshift of $z = 1.382$ but is blended with a foreground system at $z = 1.087$ (see appendix A of L. A. Boogaard et al. 2019). We adopt the redshift of the foreground system ($z = 1.087$) in this study.

For the *JWST* data, we use the v7.2 GOODS-South mosaic release, also processed with the `grizli` pipeline, with detailed analyses described in F. Valentino et al. (2023). The DJA GOODS-South mosaic combines recent *JWST* GTO and GO programs, including FRESCO (PID: GO 1895; PI: P. Oesch; P. A. Oesch et al. 2023), JEMS (PID: GO 1963; PI: C. Williams; C. C. Williams et al. 2023), JADES (PIDs: GTO 1180, GTO 1181, GTO 1210, GTO 1286, GO 1895, GO 1963, and GO 3215; PIs: D. Eisenstein & N. Luetzgendorf; M. J. Rieke et al. 2023; A. J. Bunker et al. 2024; K. N. Hainline et al. 2024; F. D'Eugenio et al. 2025; D. J. Eisenstein et al. 2025, 2026), SMILES (PID: GTO 1207; PI: G. Rieke; S. Alberts et al. 2024; G. H. Rieke et al. 2024), and MIDIS (PIDs: GTO 1283 and GO 6511; PI: G. Östlin; L. A. Boogaard et al. 2024; G. Östlin et al. 2025). In this work, we include DJA-reduced mosaics for most *JWST* wide-band filters: F090W, F115W, F150W, F200W, F277W, F356W, and F444W from NIRCam, as well as F560W from MIRI. For two MIRI wide-bands, F770W and F1000W, the DJA mosaics include only the shallower SMILES observations; therefore, we adopt the deeper mosaics[1] from the MIDIS and MIDIS-RED surveys (G. Östlin et al. 2025; P. G. Pérez-González et al. 2025; Östlin et al. in preparation) for the subsequent analyses. For the MIRI F1280W, F1500W, F1800W, F2100W, and F2550W

[1]The data are available at the Zenodo repository (J. Melinder 2025).

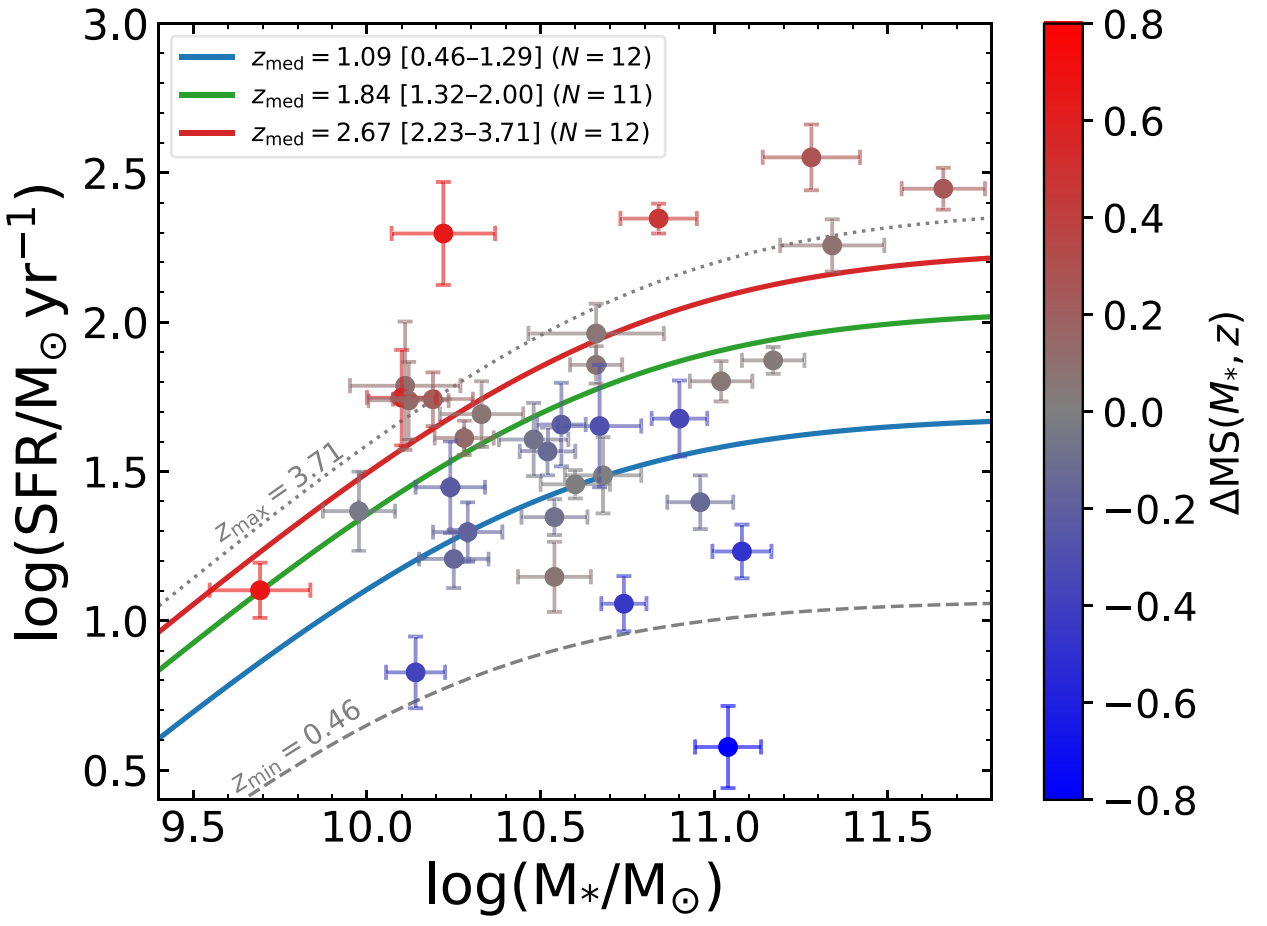

**Figure 2.** Distribution of the ASPECS galaxies on the $M_*$–SFR plane, with physical properties derived using the high-$z$ extension of `MAGPHYS` (E. da Cunha et al. 2008; E. da Cunha et al. 2015; A. J. Battisti et al. 2020). Circles represent individual galaxies, colour-coded by their offset from the SFMS (P. Popesso et al. 2023), defined as $\Delta\mathrm{MS} = \log[\mathrm{SFR}] - \log[\mathrm{SFR_{MS}}(M_*, z)]$. The SFMS at the median redshift of each of the three redshift bins is shown by the blue, green, and red lines from low to high redshift. The three bins contain approximately equal numbers of galaxies, with the number in each bin indicated in the legend. The dashed and dotted lines represent the SFMS at the lowest and highest redshifts of the sample, respectively. Overall, the ASPECS galaxies are distributed across the SFMS.

filters, we adopt the observations from SMILES (S. Alberts et al. 2024; G. H. Rieke et al. 2024).

In addition to the `SCI` extension provided by the DJA, MIDIS, and SMILES teams, we also use the `WHT` and `EXP` extensions to construct the uncertainty maps. These uncertainty maps are generated following the DJA guidelines and account for noise contributions from the sky background, instrumental effects, and Poisson statistics. Example cut-outs for one of the ASPECS galaxies are shown in Fig. 3. We also adopt the segmentation maps provided by the DJA to identify and exclude nearby sources that may contaminate our targets. For the source positions, we use the DJA catalogue positions for all galaxies in our sample, except for 1mm.C23, for which we adopt the position from the JADES catalogue. This is because the two spiral galaxies in this system are identified as a single object in the JADES catalogue, such that the reported position lies between the two galaxies.

Based on visual inspection of the *HST* and *JWST* images as well as previous studies (R. Decarli et al. 2016b; L. A. Boogaard et al. 2019, 2024), a small fraction of sources show signs of spatial blending, close companions, or ongoing interactions. Specifically, 1mm.C01 is a likely interacting galaxy pair with a companion located $\sim 1.5$ arcsec to the west at the same redshift. 1mm.C12 is located in an overdense region where several sources at different redshifts overlap in projection. 1mm.C22 is spatially blended with a foreground galaxy to the east. 1mm.C23 consists of a pair of interacting galaxies at $z \approx 1.087$. 1mm.C24 features a blue, likely foreground point source on its eastern edge, which is also masked out by the segmentation map. For most sources, contamination from neighbouring sources is effectively removed by the segmentation maps, with only 1mm.C23 remaining un-deblended due to strong spatial overlap with neighbouring sources.

## 2.3 High-resolution ALMA data

The ALMA Band 8 data used in our spatially resolved SED modelling come in part from the partially completed Cycle 8 program 2021.1.00104.S (PI: E. da Cunha). The observations were carried out in September 2022, using the C43-2 configuration, with baselines ranging from 15 to 500 m. The frequency set-up is centred at 405 GHz (740 µm in the observed frame) and covers 397–413 GHz using two tunings with four 2 GHz sidebands in Time Division Mode. The observations have seven pointings, which target 11 galaxies in the HUDF, with ten of them in our sample.

We adopt the pipeline-calibrated products and image the primary beam-corrected dust continuum using the `TCLEAN` task in `CASA` (CASA Team et al. 2022). We do not identify any emission lines in the data. Different imaging weightings were tested, and we confirm that the measured fluxes are consistent within uncertainties across weightings using a simple test based on fluxes derived with the `imfit` task. We ultimately use the cleaned images created with a robust 0.5 weighting, which provides a balance between sensitivity and angular resolution. The median root mean square (rms) of the derived maps is 373 µJy beam$^{-1}$, with a median synthesized beam size of 0.55 arcsec$\times$0.36 arcsec. The Band 8 dust continuum of the ten galaxies is shown as cyan contours in Fig. 1.

In addition to the Band 8 data, four galaxies (two of which have Band 8 coverage) in our sample also have Band 6 observations from M. Kaasinen et al. (2020). They combine low-resolution ASPECS data (PID: 2016.1.00324.L; J. González-López et al. 2020) with high-resolution 1.3 mm observations (PID: 2012.1.00173.S; J. S. Dunlop et al. 2017) and image the dust continuum with natural weighting. In this work, we re-image the dust continuum at higher angular resolution using a robust 0.5 weighting and clean down to the 2 $\sigma$ level. This results in a median beam size of 0.43 arcsec$\times$0.35 arcsec and a median rms of 26 µJy beam$^{-1}$. The dust continuum of the four galaxies is shown as magenta contours in Fig. 1. In total, there are 12 galaxies in our sample with Band 8 or Band 6 high-resolution imaging, with two (1mm.C06 and 3mm.09) having both.

## 2.4 Ancillary data

Due to the location of the ASPECS galaxies within the HUDF, they benefit from exceptional multiwavelength coverage from numerous observatories. We therefore incorporate archival data from these complementary surveys to better constrain our spatially unresolved SED modelling.

D. Elbaz et al. (2011) present *Spitzer* and *Herschel* observations covering 3.6–500 µm in the GOODS fields. In this work, we include only the deblended *Herschel* PACS 100 and 160 µm photometry. We do not use the *Spitzer* IRAC bands (3.6, 4.5, 5.8, and 8.0 µm) and the MIPS 24 µm data, as these wavelength ranges are covered by deeper *JWST* filters. The MIPS 70 µm band is also excluded due to shallow sensitivity, and the *Herschel* SPIRE bands (250, 350, and 500 µm) are avoided because of strong blending caused by their modest spatial resolution (see L. A. Boogaard et al. 2019, 2020). Although the number of bands adopted is limited, the PACS observations at 100 and 160 µm provide crucial constraints on the dust luminosity, as they lie close to the peak of dust emission.

In addition to the spatially resolved ALMA data mentioned in Section 2.3, we also include the spatially unresolved 1.2 and 3 mm dust continuum measurements from ASPECS (J. González-López et al. 2019, 2020), which help constrain the dust models in the

**Table 2.** Properties of the *HST* and *JWST* data used in this work.

| Instrument | Filter | Image ref. | Pixel scale (mas) | PSF ref. | PSF FWHM (mas) | Resolved | | | Unresolved |
|---|---|---|---|---|---|---|---|---|---|
| | | | | | | F444W_hn | F1280W_hnm | F1280W_hnma | |
| (1) | (2) | (3) | (4) | (5) | (6) | (7) | (8) | (9) | (10) |
| *HST*/ACS | F435W | DJA | 40 | Stacked | 70 | ✓ | ✓ | ✓ | ✓ |
| | F606W | DJA | 40 | Stacked | 80 | ✓ | ✓ | ✓ | ✓ |
| | F775W | DJA | 40 | Stacked | 85 | ✓ | ✓ | ✓ | ✓ |
| | F814W | DJA | 40 | Stacked | 87 | ✓ | ✓ | ✓ | ✓ |
| | F850LP | DJA | 40 | Stacked | 95 | ✓ | ✓ | ✓ | ✓ |
| *JWST*/NIRCam | F090W | DJA | 20 | STPSF | 88 | ✓ | ✓ | ✓ | ✓ |
| | F115W | DJA | 20 | STPSF | 92 | ✓ | ✓ | ✓ | ✓ |
| | F150W | DJA | 20 | STPSF | 94 | ✓ | ✓ | ✓ | ✓ |
| | F200W | DJA | 20 | STPSF | 106 | ✓ | ✓ | ✓ | ✓ |
| | F277W | DJA | 40 | STPSF | 114 | ✓ | ✓ | ✓ | ✓ |
| | F356W | DJA | 40 | STPSF | 131 | ✓ | ✓ | ✓ | ✓ |
| | F444W | DJA | 40 | STPSF | 148 | ✓ | ✓ | ✓ | ✓ |
| *JWST*/MIRI | F560W | DJA | 40 | STPSF | 207 | | ✓ | ✓ | ✓ |
| | F770W | MIDIS | 40 | STPSF | 269 | | ✓ | ✓ | ✓ |
| | F1000W | MIDIS | 40 | STPSF | 328 | | ✓ | ✓ | ✓ |
| | F1280W | SMILES | 60 | STPSF | 420 | | ✓ | ✓ | ✓ |
| | F1500W | SMILES | 60 | STPSF | 488 | | | | ✓ |
| | F1800W | SMILES | 60 | STPSF | 591 | | | | ✓ |
| | F2100W | SMILES | 60 | STPSF | 674 | | | | ✓ |
| | F2550W | SMILES | 60 | STPSF | 803 | | | | ✓ |

*Notes.*
(1) Instrument.
(2) Filter name.
(3) Image reference (DJA: Dawn *JWST* Archive; MIDIS: G. Östlin et al. 2025; P. G. Pérez-González et al. 2025; Östlin et al., in preparation.; SMILES: S. Alberts et al. 2024; G. H. Rieke et al. 2024).
(4) Pixel scale of the image. All images are reprojected onto a common pixel grid as described in Section 3.1.2.
(5) PSF reference (Stacked: empirical stacked PSF; STPSF: M. D. Perrin et al. 2012, 2014; see Section 3.1.3).
(6) PSF FWHM. After PSF matching, the target PSF is slightly broader than the original PSF owing to the suppression of high-frequency noise in Fourier space (Section 3.1.3).
(7)–(9) Spatially resolved SED-modelling configurations in which each filter is used (Section 3.2.2).
(10) All filters listed in this table are used in the spatially unresolved modelling.

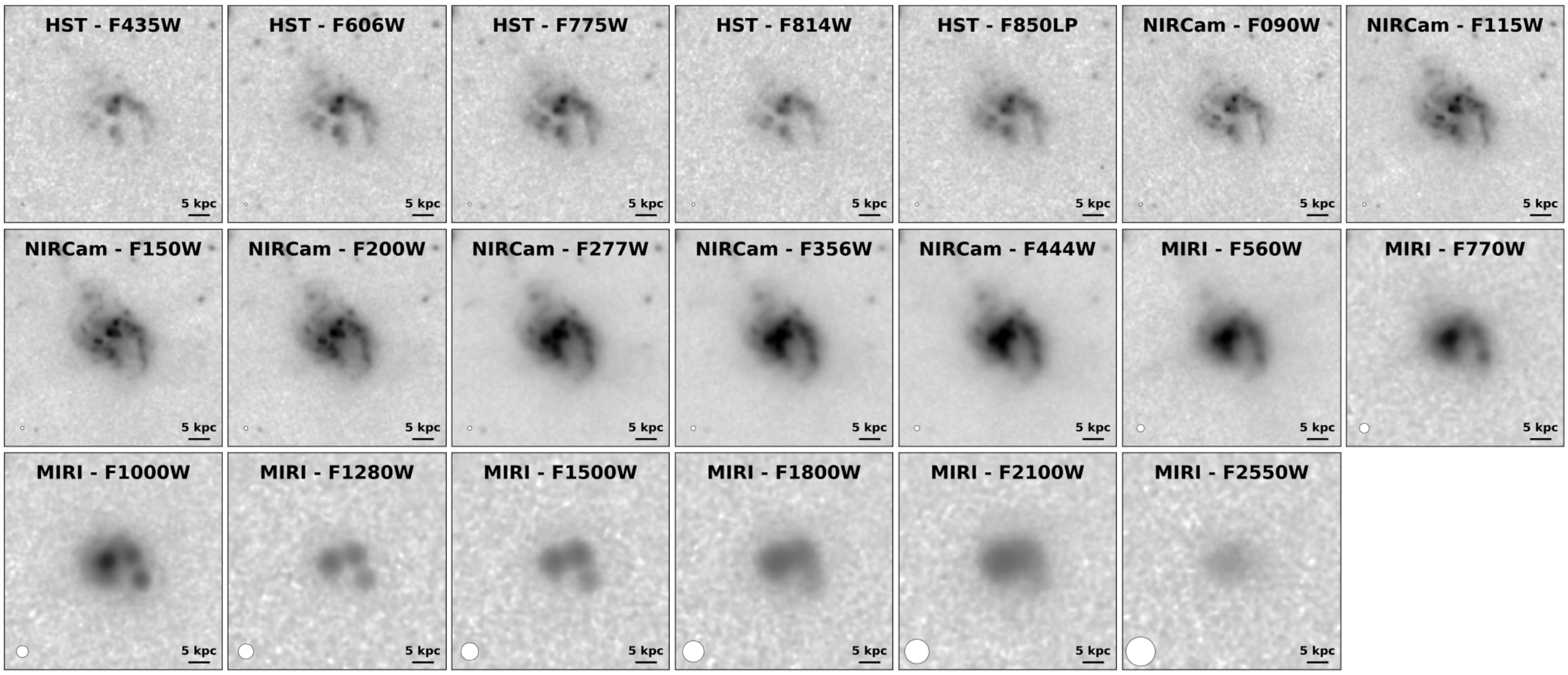

**Figure 3.** Multiwavelength 6 arcsec × 6 arcsec cut-outs of an example source, 1mm.C10 at $z = 1.997$, showing the *HST* and *JWST* filters used in the SED modelling. The PSF full width at half-maximum for each filter is indicated in the bottom-left corner of each panel, and a 5 kpc scale bar is shown at the bottom right.

Rayleigh–Jeans regime. In total, all galaxies in our sample have 1.2 mm flux measurements, and six of them additionally have 3 mm data.

## 3 ANALYSIS

In the following subsections, we outline the image processing steps before photometric extraction (Section 3.1) and describe the procedures used for the SED modelling in this work (Section 3.2).

### 3.1 Image processing

Given that the high-resolution data from *HST*, *JWST*, and ALMA are obtained with different instruments and reduced using different pipelines, we follow a series of steps to homogenize the images, ensuring consistent alignment across all data sets. We then perform aperture photometry to extract both gridded and integrated fluxes from these homogenized images. These procedures mainly follow the methodology described in J. Li et al. (2024). Below, we provide the descriptions of image processing.

#### *3.1.1 Astrometric alignment*

Most of our *HST* and *JWST* imaging data are retrieved from the DJA (F. Valentino et al. 2023), which ensures self-consistent astrometry by aligning all filters to the *Gaia* DR3 reference frame (Gaia Collaboration 2021). We also incorporate MIDIS and SMILES imaging; while MIDIS is tied to the Hubble Legacy Field catalogue (K. E. Whitaker et al. 2019; G. Östlin et al. 2025), and SMILES is aligned with the JADES reference catalogue within $\sim 0.01$–$0.02$ arcsec accuracy (S. Alberts et al. 2024; B. D. Johnson et al. 2026), all these catalogues are ultimately anchored to the same *Gaia* absolute reference system. Furthermore, we find no systematic offsets between the ALMA detections and their *JWST* counterparts that exceed the nominal ALMA positional accuracy ($\sim \theta_{\rm beam}$/SNR).[2] Notably, these alignment offsets are significantly smaller than our target resolutions (0.18 and 0.50 arcsec; see Section 3.1.3) and aperture grid sizes (0.04 arcsec; see Section 3.1.5). Therefore, we expect that uncertainties in the astrometric alignment are unlikely to significantly impact our spatially resolved analysis, given the spatial scales employed in this work.

#### *3.1.2 Reprojection and cut-out*

We first aligned all multiwavelength mosaics onto a common pixel grid, using the F444W mosaic as the reference image. The reprojection was performed using the `reproject_adaptive` function in the ASTROPY package (Astropy Collaboration 2022), with flux conservation enabled. This produces a homogenized pixel grid for all images across filters, with a uniform pixel size of 40 mas. We then create 30 arcsec $\times$ 30 arcsec cut-outs for each galaxy based on the coordinates listed in Table 1 from the reprojected mosaics of each filter.

#### *3.1.3 Point spread function matching*

Next, we ensure that all images are affected by the same optical diffraction patterns, preventing artificial colours caused by differences in the point spread functions (PSFs) of various filters. For each filter, the cut-out image is convolved with a corresponding kernel using the `convolve` function in the ASTROPY package. The kernel is filter-dependent and derived from the ratio of the target PSF (F444W or F1280W in this work; see Section 3.2) to the PSF of the specific filter in the Fourier space. These steps are implemented using the `create_matching_kernel` function in the `photutils` package (L. Bradley et al. 2024), with a 2D `CosineBellWindow` function applied to suppress high-frequency noise. The `CosineBellWindow` has a single free parameter, $\alpha$, which controls the extent of filtering: lower $\alpha$ values produce stronger filtering and a wider derived kernel, and vice versa. Based on our testing, we determine optimal $\alpha$ values of 0.70 and 0.25 for matching to the F444W and F1280W PSFs, respectively. These values provide the weakest filtering that effectively suppresses high-frequency noise while preserving as much angular resolution as possible, and are therefore adopted in the following analyses. The final PSF-matched angular resolutions are 0.18 and 0.50 arcsec for F444W and F1280W, respectively. Fig. 4 (a) and (b) illustrates the F090W images before and after PSF matching to the F444W resolution.

The PSFs for the NIRCam and MIRI filters are simulated using the STPSF package (M. D. Perrin et al. 2012, 2014), which models the PSF by accounting for the optical path differences and detector response. We generate the PSFs with an oversampling factor of three and rotate them to match the position angle (PA) of the observed images. For NIRCam filters, we adopt the `PA_APER` from the JADES observations, as the JADES images are the primary contributors to the DJA mosaics. For MIRI F560W, F770W, and F1000W, the `PA_APER` is taken from the MIDIS mosaic headers, while for MIRI F1280W, we use the value from the SMILES mosaic header. After rotation, we visually verify that the PSFs are aligned with the point sources in the field.

The PSFs for the *HST* filters used in this study (F435W, F606W, F775W, F814W, and F850LP) are empirical, derived by stacking point sources. A summary of the PSF properties for the *HST* and *JWST* filters is presented in Table 2. For the ALMA data, we follow the approach of J. Li et al. (2024) and generate the PSFs using 2D elliptical Gaussian models, with the major axis, minor axis, and PA determined by the `BMAJ`, `BMIN`, and `PA` keywords in the ALMA image headers, respectively.

#### *3.1.4 Segmentation*

To minimize contamination from nearby sources, we retain only the pixels associated with our targets, based on the segmentation maps provided by DJA. The sources are identified using SEP (K. Barbary 2016), a Python implementation of SExtractor (E. Bertin & S. Arnouts 1996), applied to a noise-weighted combination of the NIRCam long-wavelength filters (typically F277W, F356W, and F444W). Further details on the source detection and segmentation can be found in F. Valentino et al. (2023).

#### *3.1.5 Pixel binning and photometry*

Lastly, we construct an aperture grid within a circular region[3] that encompasses all flux from the targets in every filter, as illustrated in Fig. 4 (d), to extract photometry for the spatially resolved SED modelling. Each aperture element consists of a 4 $\times$ 4 block of 40 mas pixels (corresponding to 0.16 arcsec × 0.16 arcsec) and serves as an individual element (hereafter: spatial bin) in our spatially resolved modelling. For the spatially unresolved modelling, the photometry is obtained by summing all pixels within the same segmentation maps described in Section 3.1.4.

[2] More details in Chapter 10.5.2 of ALMA Cycle 13 Technical Handbook.

[3] With a radius of 2.88 arcsec for $z > 1$ sources and 4.8 arcsec for $z \leq 1$ sources, which are large enough to cover all the targets by visual inspection.

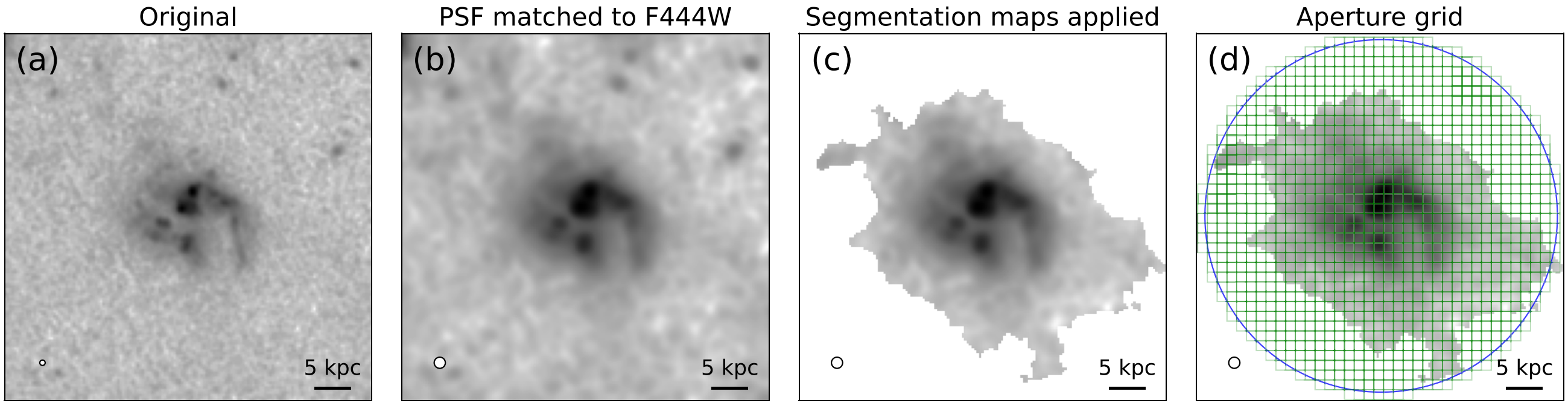

**Figure 4.** Example cut-outs of the NIRCam F090W filter of 1mm.C10, illustrating the different image processing steps. The angular resolution is indicated in the bottom left corner of each panel, and a 5 kpc scale bar is shown at the bottom right. (a) Original observed image. (b) Image convolved with a kernel to match the F444W PSF. (c) Image with the segmentation map applied. (d) Aperture grid used for photometry measurements.

We note that the PSF-matching process mentioned in Section 3.1.3 introduces additional correlations between neighbouring pixels, so the covariance terms of the error propagation formula must be accounted for when binning the uncertainty maps of the *HST*, *JWST*, and ALMA observations. To address this, we follow the idea presented in G. Östlin et al. (2025) (see their section 3.1) and calculate a correlated-to-uncorrelated correction factor, $R_{\rm cor}$, and multiply it by the uncertainty estimated under the assumption that the binned pixels are fully independent (i.e. the square root of the sum of the squared uncertainties of all pixels within the bin).

To determine $R_{\rm cor}$, we randomly place our 4 pix × 4 pix (0.16 arcsec × 0.16 arcsec) spatial bin aperture on the 2.5 arcmin × 2.5 arcmin PSF-matched background science images, where all science targets are masked. This background cut-out is centred at ($3^{\rm h}32^{\rm m}39\overset{\rm s}{.}45$, $-27^{\rm d}47^{\rm m}10\overset{\rm s}{.}00$), roughly the centre of our galaxy sample. This background sampling process is repeated $10^6$ times with a $2\sigma$-clipping applied, and the root-mean-square (rms) value is taken as the effective correlated uncertainty ($\sigma_{\rm cor}$). We then repeat the procedure using a 1 pix × 1 pix (0.04 arcsec × 0.04 arcsec) aperture to derive the pixel uncertainty ($\sigma_{\rm pix}$). The correlation factor is then computed as

$$R_{\rm cor} = \frac{\sigma_{\rm cor}}{\sigma_{\rm pix} \times \sqrt{N_{\rm bin}}}, \quad (1)$$

where $N_{\rm bin}$ is the number of pixels in the spatial bin ($N = 16$ in our $4 \times 4$ pixel binning case). In principle, $R_{\rm cor} = 1$ for fully independent pixels, and $R_{\rm cor} = \sqrt{N_{\rm bin}}$ for fully correlated pixels. From our measurements, $R_{\rm cor}$ varies slightly between filters, with an average value of $\sim 3.5$ for PSF-matching to F444W and close to 4 for F1280W.

## 3.2 SED modelling

### 3.2.1 MAGPHYS

To derive the physical properties of our galaxy sample, both in spatially resolved and unresolved schemes, we employ the high-$z$ extension v2 of the MAGPHYS code[4] (E. da Cunha et al. 2008; E. da Cunha et al. 2015; A. J. Battisti et al. 2020) to model the multiwavelength data. In brief, MAGPHYS models the UV-to-radio SEDs based on energy balance and derives physical properties via Bayesian analysis. The high-$z$ extension v2 includes the 2175 Å bump in the dust attenuation curve and expands the parameter space [e.g. star formation history (SFH) and dust opacity depth] to better accommodate high-$z$ galaxies.

Stellar emission is modelled using the stellar population synthesis models of G. Bruzual & S. Charlot (2003), adopting a G. Chabrier (2003) IMF and metallicities ranging from 0.2 to 2 solar metallicity ($Z_\odot$). The SFH is parametrized as a continuous delayed exponential function, with random bursts allowed within the past 2 Gyr prior to the epoch of observation. Dust attenuation follows S. Charlot & S. M. Fall (2000), separating attenuation into birth cloud and diffuse interstellar medium (ISM) components with different strengths. The total energy absorbed in the optical and NIR is then re-emitted in the MIR and FIR, modelled using a polycyclic aromatic hydrocarbon (PAH) template plus a set of greybody functions.

MAGPHYS constructs a library of SED templates by sampling the parameter space of different adopted models and computes both the predicted fluxes in each filter and the corresponding physical properties. The observed fluxes are compared to the predicted fluxes of all models via a $\chi^2$ goodness-of-fit metric for each model. The probability density function of each physical property is then derived by marginalizing the probability in each model by $P \propto \exp(-\chi^2/2)$. In this work, we adopt the median values along with the 16$^{\rm th}$ and 84$^{\rm th}$ percentiles of the PDFs for our analysis.

### 3.2.2 Modelling set-ups

In this study, we perform both spatially resolved and spatially unresolved SED modelling. Both types of modelling are implemented independently. For the spatially unresolved modelling, we use the *HST*, *JWST*, *Herschel*, and ALMA data introduced in Section 2. For the spatially resolved modelling, we adopt three primary configurations alongside two auxiliary configurations, each offering different wavelength coverages. For simplicity, the spatially resolved configurations are named according to the PSF-matching target filter and the instruments included in the modelling, comprising three primary configurations (the first three listed below) and two auxiliary configurations (the last two listed below), as summarized below with the number of modelled galaxies presented in brackets:

(i) **F444W_hn** ($N = 35$): *HST*/ACS + *JWST*/NIRCam
(ii) **F1280W_hnm** ($N = 35$): *HST*/ACS + *JWST*/NIRCam + *JWST*/MIRI (up to F1280W)
(iii) **F1280W_hnma** ($N = 12$): *HST*/ACS + *JWST*/NIRCam + *JWST*/MIRI (up to F1280W) + ALMA (Band 8 and/or Band 6, if available)
(iv) **F1280W_hn** ($N = 35$): *HST*/ACS + *JWST*/NIRCam
(v) **F1280W_hna** ($N = 12$): *HST*/ACS + *JWST*/NIRCam + ALMA (Band 6 and/or Band 8, if available)

[4] https://www.iap.fr/magphys/

A summary of the *HST* and *JWST* data used in the primary configurations is presented in Table 2.

The motivation for defining these primary configurations (F444W_hn, F1280W_hnm, and F1280W_hnma) is to systematically investigate the impact of progressively expanding the wavelength coverage into longer wavelength regimes on spatially resolved SED fitting. As highlighted in Section 1, most spatially resolved SED studies at high redshift have been restricted to the rest-frame optical. Our baseline configuration, F444W_hn, covers up to F444W ($\sim 1\,\mu$m in the rest frame), fully sampling the rest-frame optical spectrum. To extend the coverage into the rest-frame NIR and MIR regimes, F1280W_hnm incorporates MIRI photometry up to F1280W. We specifically adopt F1280W as the PSF-matching target because it balances extended MIR coverage with minimal degradation of spatial resolution, matching closely with the angular resolution of our ALMA data set. Finally, F1280W_hnma incorporates FIR cold dust continuum constraints from ALMA, allowing us to directly evaluate how the inclusion of spatially resolved FIR dust observations affects the derived stellar and ISM parameters.

In addition, the two auxiliary configurations (F1280W_hn and F1280W_hna) allow us to conduct further investigations. To separate the effects of wavelength coverage and spatial resolution, we introduce the auxiliary F1280W_hn configuration, with a discussion presented in Section 5.1. We also introduce the auxiliary F1280W_hna configuration to isolate the contributions of the MIR and FIR data, with a detailed comparison presented in Section 5.2. Throughout the remainder of this paper, we focus our analyses on the first three primary configurations (F444W_hn, F1280W_hnm, and F1280W_hnma).

In our spatially resolved modelling, we only include bins with detected emission at $> 1\sigma$ in more than three NIRCam filters to avoid contamination from weakly constrained spatial elements (i.e. those with $> 1\sigma$ detections in fewer than three filters). We also tested a more stringent $3\sigma$ threshold and found that the derived properties remain virtually unchanged, consistent with the findings of I. Smail et al. (2023). Bins with non-detections are incorporated into MAGPHYS using $3\sigma$ upper limits. By default, MAGPHYS imposes a minimum 10 per cent flux uncertainty to prevent overfitting. However, given the well-sampled SED coverage in our data, this threshold can be insufficient in *S*/*N* bins, preventing MAGPHYS from properly sampling the full model parameter space and leading to artificially narrow likelihood distributions. To mitigate this, we adopt a higher minimum flux uncertainty of 20 per cent, which provides a more realistic representation of systematic uncertainties and ensures a more robust exploration of the model space.

### *3.2.3 MAGPHYS model restriction*

As we will show in Section 4.2.3 and discuss further in Section 5.3, the total dust luminosities obtained from the spatially resolved modelling ($L_{\rm dust,resv}$), computed by summing over all spatial bins, are systematically higher than those derived from the spatially unresolved modelling ($L_{\rm dust,unresv}$) that includes MIR and FIR constraints. This behaviour is seen in all three modelling configurations. The discrepancy arises because, in the absence of spatially resolved FIR data, the FIR emission is primarily driven by the priors in MAGPHYS, which may not be appropriate for every galaxy or for all spatial regions. The priors in MAGPHYS are defined via the distribution of a pre-generated set of models, meaning that modifying the priors relies on changing the model density. To address this, we use the observed unresolved colours to impose additional constraints on the MAGPHYS model library, excluding models that deviate significantly from the observed unresolved colours and thereby narrowing the parameter space explored in the spatially resolved modelling. Throughout this work, we therefore distinguish between two MAGPHYS model configurations:

(i) The **default models**, which include the complete default model library.
(ii) The **restricted models**, in which models broadly consistent with the observed spatially unresolved colours are adopted.

The model restricting procedure is implemented by retaining only those MAGPHYS models whose colours are broadly consistent with those measured from the spatially unresolved photometry. In this way, the spatially unresolved photometry is used to inform and constrain the spatially resolved SED fitting. Specifically, we adopt the colour indices F560W/F2550W, F560W/PACS100, and F560W/BAND6, which probe the PAH emission, the peak of the dust SED, and the Rayleigh–Jeans regime, respectively. Only models with colours within $\pm 0.65$ dex of the observed values are retained. This adopted tolerance of $\pm 0.65$ dex is based on the standard deviation of the F560W/BAND6 colour measured across all spatial bins of galaxies with spatially resolved ALMA Band 6 detections. The best-fitting SEDs obtained using the restricted models are shown in Fig. 5, and the possible impact of the selected tolerance is discussed in Section 5.3.

We further illustrate the effect of the model restriction on the best-fitting SEDs. Fig. 6 presents the results for a representative galaxy exhibiting one of the most pronounced improvements, comparing spatially unresolved modelling with the three spatially resolved configurations. In the MIR and FIR regimes, the SEDs derived from the restricted models are systematically closer to both the spatially unresolved photometric data points and the corresponding best-fitting unresolved SED than those obtained using the default model library.

Fig. 6 (left panel) also highlights the role of MIRI and ALMA observations in constraining the SED when using the default model library. In the F444W_hn configuration, the best-fitting SED overestimates the emission at wavelengths $\lambda_{\rm obs} \gtrsim 10\,\mu$m. Including MIRI data in the F1280W_hnm configuration substantially reduces this discrepancy, bringing the SED closer to the spatially unresolved solution. When ALMA data are further incorporated in the F1280W_hnma configuration, the agreement improves further, particularly in the Rayleigh–Jeans regime. These improvements in the SED fits are reflected in more reliable estimates of dust-related physical properties, as discussed in Sections 4.2.3.

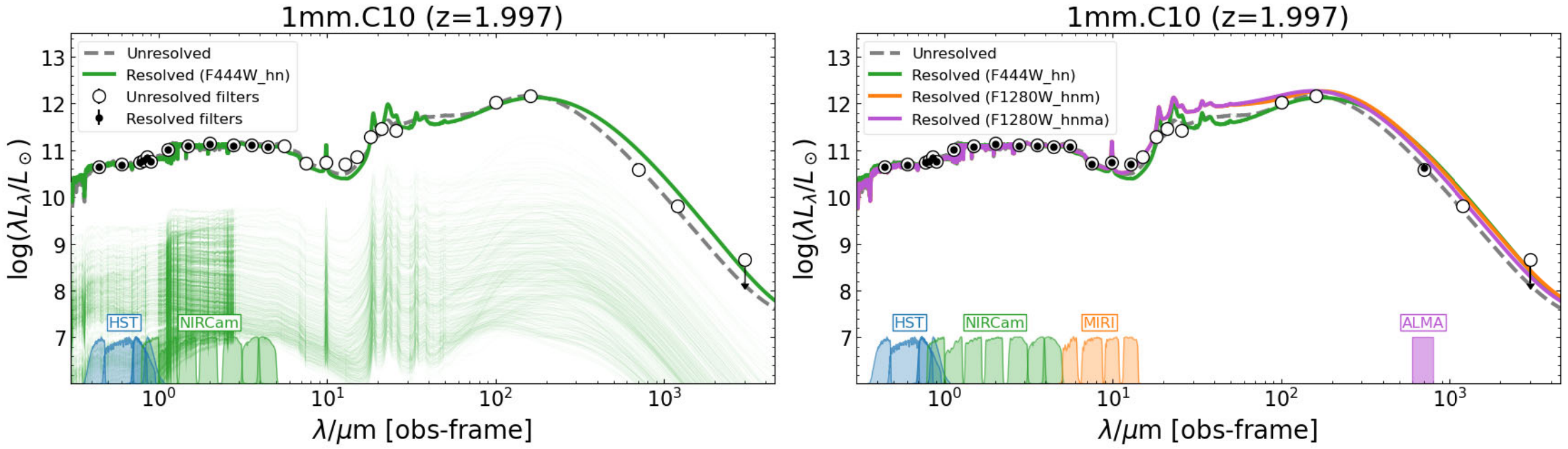

**Figure 5.** Photometry and best-fitting SEDs from the restricted-model MAGPHYS fittings for the example galaxy 1mm.C10. Back-filled circles indicate the summed photometry used in the spatially resolved modelling, while open circles show the photometry used in the spatially unresolved modelling. The grey-dashed curve represents the best-fitting SED from the unresolved modelling. The transmission curves of the filters used in the resolved fittings are shown at the bottom of each panel.Left**:** The thin green curves show the best-fitting SEDs of individual spatial bins, while the thick green curve represents the SED obtained by summing the curves of all the spatial bins. Right: Comparison of the summed SEDs from the three restricted-model spatially resolved set-ups described in Section 3.2.2: F444W_hn (green; using data from *HST*/ACS and *JWST*/NIRCam), F1280W_hnm (orange; using data from *HST*/ACS, *JWST*/NIRCam, and *JWST*/MIRI up to F1280W), and F1280W_hnma (purple; using data from *HST*/ACS, *JWST*/NIRCam, *JWST*/MIRI up to F1280W, and ALMA band 8). The black-filled circles are shown for the F1280W_hnma configuration as a representation. The best-fitting SED curves from the spatially resolved and unresolved fittings show good agreement across the wavelength range covered by the resolved filters.

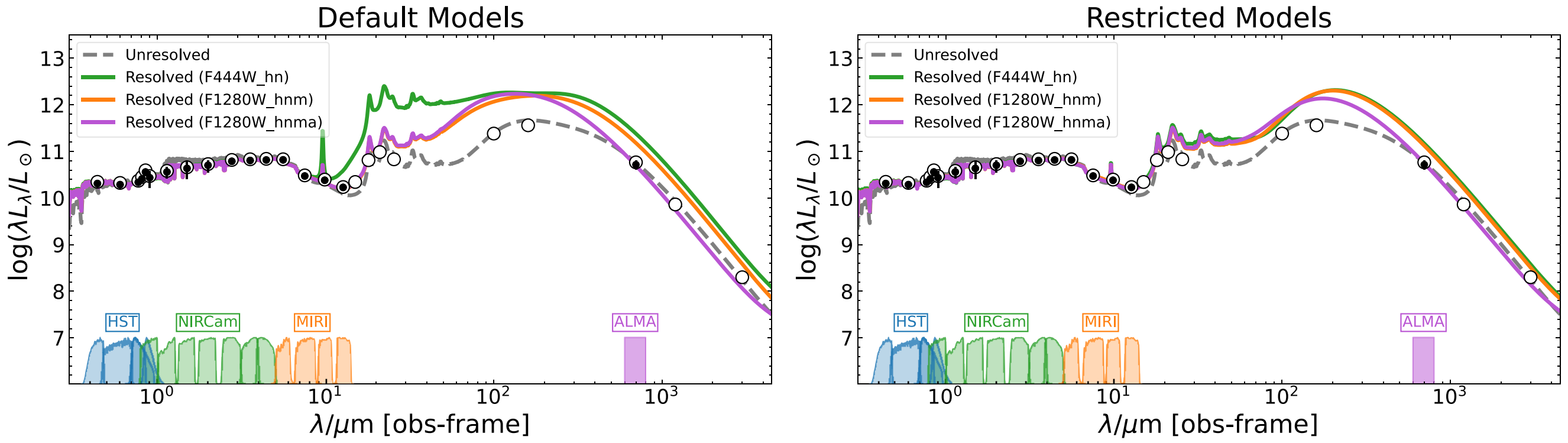

**Figure 6.** Resolved SED modelling results for the source 1mm.C02, which exhibits one of the larger differences between the default (left panel) and restricted (right panel) modelling. Figure symbols are the same as those in Fig. 5, with the black-filled circles representing the resolved photometric data points for the F1280W_hnma configuration as a representation. The inclusion of MIRI and ALMA data improves the inferred resolved SEDs in the MIR and Rayleigh–Jeans regimes in the default model. In the restricted model, the refinement from the inclusion of MIRI and ALMA data is less prominent than in the default model, as the model restrictions already provide constraints to first order.

## 4 RESULTS

### 4.1 Unresolved SED variations and composite SED

In this section, we examine the unresolved SEDs of our cosmic noon MS galaxies to assess their intrinsic variation and evaluate how their median SED compares with local galaxy populations.

In the top-left panel of Fig. 7, we show the spatially unresolved best-fitting SEDs of all ASPECS galaxies, normalized to the sample median $\log(L_{\mathrm{IR}}/\mathrm{L}_{\odot}) = 11.6$. Each SED is colour-coded by redshift. We also construct a composite SED, defined as the median at each wavelength grid point, for the full sample and three redshift bins containing approximately equal numbers of galaxies. For each redshift bin, the composite SED is normalized to the median $\log(L_{\mathrm{IR}}/\mathrm{L}_{\odot})$ of that bin, which is 11.3, 11.9, and 12.1 from the lowest- to highest-redshift bin, respectively. We find that the composite SEDs show stronger UV/optical attenuation and a shorter-wavelength dust emission peak towards higher redshifts. This redshift evolution is primarily driven by the increasing $L_{\mathrm{IR}}$ of the galaxies, as higher-$L_{\mathrm{IR}}$ sources tend to exhibit stronger dust attenuation and warmer dust emission, consistent with trends observed in other galaxy populations (H. J. Zahid et al. 2013; H. Domínguez Sánchez et al. 2014; E. da Cunha et al. 2015; U. Dudzevičiūtė et al. 2020). In the bottom-left panel, we show the fraction of the sample with observational constraints at each wavelength, determined from the filter transmission curves and the redshift of each galaxy. A value of 100 per cent indicates that all sources have observational constraints at that wavelength.

In the top-right panel of Fig. 7, we compare the composite SED of our cosmic noon MS galaxies with three local galaxy templates from M. Polletta et al. (2007): normal star-forming galaxies (represented by the Sd template), starburst galaxies (represented by M82), and ultraluminous infrared galaxies (ULIRGs; represented by Arp 220). All curves are normalized at rest-frame 350 μm, corresponding approximately to the average rest-frame selection wavelength of the ASPECS sample. We find that the composite SED closely resembles that of M82, suggesting that $z \approx$ 1-3 MS galaxies share broadly similar dust attenuation and re-emission properties with local starburst galaxies. The apparent deficit around $\sim 20$ µm relative to M82 is more uncertain, as this wavelength range is poorly constrained for our sample, which is indicated by the constraint percentage shown in the lower panel. Future FIR facilities, such as PRIMA (A. Moullet et al. 2025), capable of probing this wavelength regime will therefore be valuable for better constraining the warm-dust emission of these cosmic noon galaxies. Compared to local normal star-forming galaxies, cosmic noon MS galaxies exhibit substantially stronger infrared emission and warmer dust temperatures, as evidenced by their dust emission peak at $\sim$ 60–70 µm. In contrast, although they do not show the extreme UV-to-IR attenuation characteristic of local ULIRGs such as Arp 220, they display similar emission in the Rayleigh–Jeans regime, suggesting comparable contributions from cold dust to the total infrared luminosity. Overall, local starburst galaxies such as M82 serve as appropriate empirical analogues for $z \approx$ 1-3 MS galaxies in terms of their integrated SED shapes.

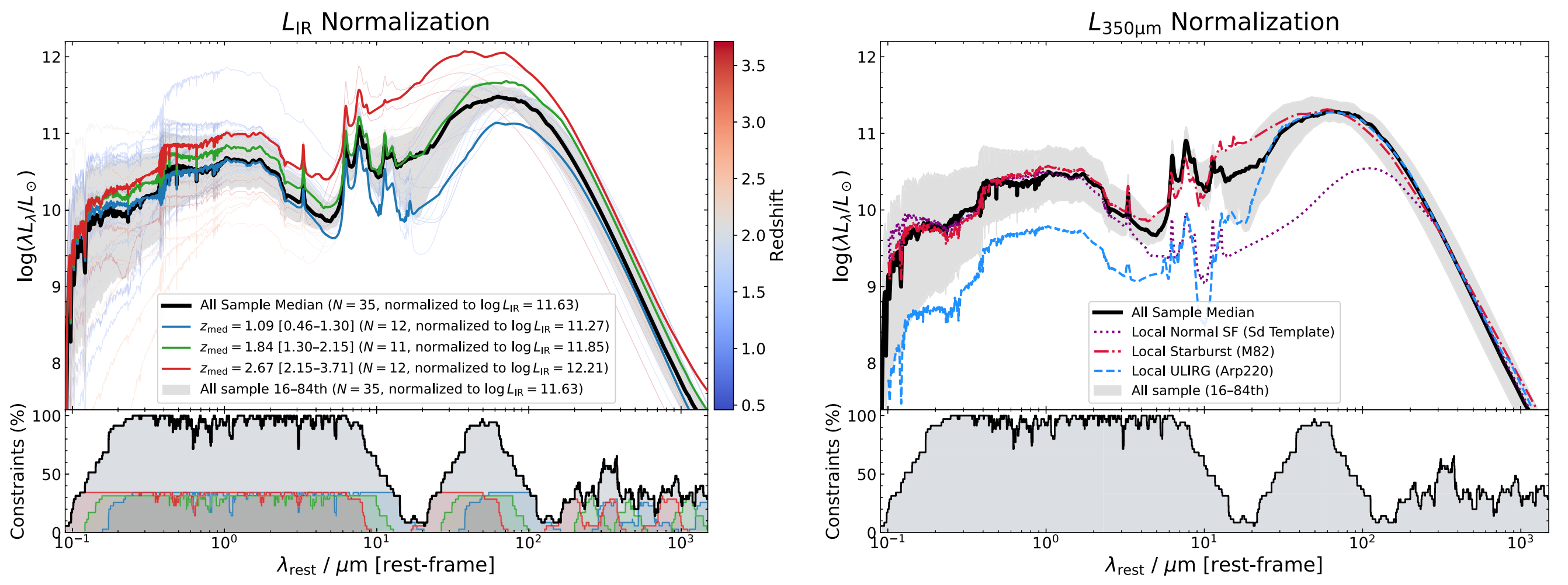

**Figure 7.** SED comparisons across the ASPECS sample and local galaxy populations, with the fraction of the ASPECS sample having observational constraints at each wavelength determined from the filter transmission curves and redshifts.Left: Spatially unresolved best-fitting SEDs of individual ASPECS galaxies (thin curves), colour-coded by redshift and normalized to the sample median $\log(L_{IR}/L_\odot) = 11.6$. The composite SEDs, defined as the median at each wavelength grid point, are shown for the full sample and three redshift bins ($z = 0.46$–$1.30$, $z = 1.30$–$2.15$, and $z = 2.15$–$3.71$) by the black, blue, green, and red lines, respectively. The grey-shaded band represents the $16^{th}$–$84^{th}$ percentile range of the composite SED for the full sample. The composite SED of each redshift bin is normalized to its average $L_{IR}$, with the corresponding values indicated in the legend. The higher-redshift bins show stronger UV/optical attenuation and warmer dust emission, primarily driven by the increasing $L_{IR}$ with redshift. Right: Comparison of the cosmic noon composite SED with local benchmark galaxy templates from M. Polletta et al. (2007): normal star-forming galaxies (Sd template; purple dotted line), starburst galaxies (M82; red dash–dotted line), and ULIRGs (Arp,220; blue dashed line). All curves are normalized at the rest-frame 350 µm, approximately corresponding to the rest-frame wavelength at which the ASPECS galaxies were selected. Overall, the $z \approx$ 1-3 MS galaxies in our sample show integrated SED shapes similar to those of local starburst galaxies such as M82. The tabulated composite SEDs of the ASPECS sample are provided in the supplementary files.

## 4.2 Physical properties

In the following subsections, we present the SED-derived physical properties, including stellar mass ($M_*$), 100 Myr-averaged SFR, specific star formation rate (sSFR), dust luminosity ($L_{dust}$), and dust mass ($M_{dust}$), derived from both the spatially unresolved and spatially resolved modelling (F444W_hn, F1280W_hnm, and F1280W_hnma). We then compare the results obtained from our resolved and unresolved modelling approaches and discuss possible causes of any discrepancies.

### 4.2.1 *Spatially unresolved*

We present the physical properties derived from the spatially unresolved modelling in Table A1. When compared with the previous estimates from M. Aravena et al. (2020), who also used MAGPHYS, we find good agreement within $\sim 0.1$ dex among all the derived physical properties. Minor discrepancies likely arise from differences in the adopted redshifts and the MIR photometry. In this study, we use the spectroscopic redshift measurements presented in L. A. Boogaard et al. (2024) and replace the lower-resolution Spitzer measurements, potentially affected by blending with nearby sources, with higher-resolution and more sensitive *JWST* data. Another possible reason for the discrepancy stems from the different MAGPHYS versions used: M. Aravena et al. (2020) employed the high-$z$ v1 code, whereas we use the updated high-$z$ v2 one. The latter allows for the presence of the 2175 Å dust attenuation bump, parametrized by the bump strength $E_b^{'}$, which was not included in the earlier version. Recent studies at higher redshift (e.g. R. Fisher et al. 2025) have shown that omitting this bump can significantly affect the derived stellar masses, though we do not find a clear correlation between $E_b^{'}$ and the $M_*$ offsets in our sample. Despite these small differences, our SED-derived physical properties remain consistent with those from M. Aravena et al. (2020), indicating that the overall understanding of the ASPECS galaxies is unchanged.

In addition to the MAGPHYS-derived quantities, we also estimate the global molecular gas mass ($M_{mol}$) using the dust mass ($M_{dust}$) obtained from the spatially unresolved SED modelling. We adopt a dust mass absorption coefficient normalized to $\kappa_{850\,\mu m} = 0.4699\,g^{-1}\,cm^2$ (B. T. Draine & A. Li 2007; B. T. Draine et al. 2014), and a molecular gas-to-dust ratio ($\delta_{GDR}$) of 200, consistent with the previous ASPECS study (M. Aravena et al. 2020).

**Table 3.** Integrated physical properties derived from the optimal spatially resolved configuration from the model restriction method.

| ID (1) | $\log_{10}(M_*/\mathrm{M}_\odot)$ (2) | $\log_{10}(\mathrm{SFR}/\mathrm{M}_\odot\,\mathrm{yr}^{-1})$ (3) | $\log_{10}(L_{\mathrm{dust}}/\mathrm{L}_\odot)$ (4) | $\log_{10}(M_{\mathrm{dust}}/\mathrm{M}_\odot)$ (5) | Configuration (6) |
|---|---|---|---|---|---|
| 1mm.C01 | $10.31 \pm 0.29$ | $2.14 \pm 0.20$ | $12.29 \pm 0.22$ | $8.21 \pm 0.49$ | F444W_hn |
| 1mm.C02 | $10.85 \pm 0.14$ | $1.97 \pm 0.19$ | $12.04 \pm 0.19$ | $8.21 \pm 0.42$ | F1280W_hnma |
| 1mm.C03 | $11.22 \pm 0.27$ | $1.93 \pm 0.30$ | $12.11 \pm 0.21$ | $8.20 \pm 0.48$ | F1280W_hnma |
| 1mm.C04 | $10.75 \pm 0.30$ | $1.81 \pm 0.32$ | $11.93 \pm 0.30$ | $8.09 \pm 0.49$ | F444W_hn |
| 1mm.C05 | $11.37 \pm 0.31$ | $1.78 \pm 0.31$ | $12.01 \pm 0.25$ | $8.25 \pm 0.49$ | F444W_hn |
| 1mm.C06 | $11.25 \pm 0.35$ | $2.17 \pm 0.26$ | $12.32 \pm 0.26$ | $8.53 \pm 0.48$ | F444W_hn |
| 1mm.C07[a] | $10.88 \pm 0.26$ | $1.76 \pm 0.24$ | $11.86 \pm 0.24$ | $7.98 \pm 0.48$ | F444W_hn |
| 1mm.C08[a] | $11.36 \pm 0.24$ | $2.62 \pm 0.18$ | $12.70 \pm 0.20$ | $8.15 \pm 0.24$ | F444W_hn |
| 1mm.C09 | $10.02 \pm 0.27$ | $1.47 \pm 0.28$ | $11.52 \pm 0.31$ | $7.62 \pm 0.48$ | F444W_hn |
| 1mm.C10 | $10.87 \pm 0.20$ | $2.37 \pm 0.22$ | $12.41 \pm 0.24$ | $8.27 \pm 0.44$ | F1280W_hnma |
| 1mm.C11 | $10.65 \pm 0.24$ | $1.48 \pm 0.26$ | $11.61 \pm 0.25$ | $7.88 \pm 0.45$ | F444W_hn |
| 1mm.C12 | $10.19 \pm 0.15$ | $1.51 \pm 0.20$ | $11.57 \pm 0.20$ | $7.66 \pm 0.47$ | F1280W_hnm |
| 1mm.C13 | $11.06 \pm 0.31$ | $1.30 \pm 0.38$ | $11.63 \pm 0.31$ | $7.79 \pm 0.48$ | F444W_hn |
| 1mm.C14a | $10.44 \pm 0.28$ | $1.41 \pm 0.17$ | $11.54 \pm 0.18$ | $7.51 \pm 0.48$ | F444W_hn |
| 1mm.C14b | $10.26 \pm 0.17$ | $1.55 \pm 0.23$ | $11.55 \pm 0.26$ | $7.60 \pm 0.46$ | F1280W_hnm |
| 1mm.C15 | $11.12 \pm 0.24$ | $1.17 \pm 0.34$ | $11.52 \pm 0.26$ | $7.73 \pm 0.49$ | F444W_hn |
| 1mm.C16 | $10.58 \pm 0.13$ | $1.55 \pm 0.20$ | $11.56 \pm 0.21$ | $7.74 \pm 0.47$ | F1280W_hnm |
| 1mm.C17 | $10.57 \pm 0.22$ | $1.58 \pm 0.32$ | $11.63 \pm 0.32$ | $7.71 \pm 0.46$ | F444W_hn |
| 1mm.C18 | $10.48 \pm 0.14$ | $1.77 \pm 0.20$ | $11.77 \pm 0.26$ | $7.79 \pm 0.46$ | F1280W_hnm |
| 1mm.C19 | $10.57 \pm 0.25$ | $1.73 \pm 0.22$ | $11.76 \pm 0.24$ | $7.68 \pm 0.41$ | F444W_hn |
| 1mm.C20 | $10.77 \pm 0.17$ | $1.09 \pm 0.34$ | $11.30 \pm 0.24$ | $7.56 \pm 0.50$ | F1280W_hnm |
| 1mm.C21 | $10.26 \pm 0.21$ | $1.33 \pm 0.27$ | $11.34 \pm 0.29$ | $7.39 \pm 0.46$ | F444W_hn |
| 1mm.C22 | $10.28 \pm 0.16$ | $1.40 \pm 0.22$ | $11.44 \pm 0.23$ | $7.47 \pm 0.46$ | F444W_hn |
| 1mm.C23 | $10.72 \pm 0.14$ | $1.51 \pm 0.22$ | $11.45 \pm 0.30$ | $7.53 \pm 0.60$ | F1280W_hnm |
| 1mm.C24 | $10.07 \pm 0.19$ | $1.68 \pm 0.25$ | $11.70 \pm 0.29$ | $7.71 \pm 0.47$ | F1280W_hnm |
| 1mm.C25 | $10.60 \pm 0.30$ | $1.37 \pm 0.36$ | $11.49 \pm 0.30$ | $7.64 \pm 0.49$ | F444W_hn |
| 1mm.C26 | $10.28 \pm 0.33$ | $1.20 \pm 0.28$ | $11.29 \pm 0.25$ | $7.44 \pm 0.48$ | F444W_hn |
| 1mm.C28 | $10.57 \pm 0.11$ | $1.09 \pm 0.26$ | $11.09 \pm 0.17$ | $7.50 \pm 0.36$ | F1280W_hnm |
| 1mm.C30[a] | $9.64 \pm 0.18$ | $1.17 \pm 0.28$ | $11.18 \pm 0.29$ | $7.00 \pm 0.38$ | F444W_hn |
| 1mm.C31 | $9.93 \pm 0.16$ | $1.46 \pm 0.19$ | $11.37 \pm 0.27$ | $7.34 \pm 0.46$ | F444W_hn |
| 1mm.C32 | $10.24 \pm 0.16$ | $0.40 \pm 0.27$ | $10.68 \pm 0.14$ | $7.21 \pm 0.50$ | F1280W_hnm |
| 1mm.C33 | $11.07 \pm 0.14$ | $0.55 \pm 0.35$ | $10.87 \pm 0.22$ | $7.33 \pm 0.51$ | F1280W_hnm |
| 3mm.09[a] | $11.20 \pm 0.36$ | $2.16 \pm 0.24$ | $12.29 \pm 0.24$ | $8.42 \pm 0.50$ | F444W_hn |
| 3mm.11 | $10.12 \pm 0.18$ | $0.94 \pm 0.42$ | $10.97 \pm 0.38$ | $7.09 \pm 0.63$ | F444W_hn |
| 3mm.16 | $10.26 \pm 0.14$ | $1.27 \pm 0.24$ | $11.28 \pm 0.25$ | $7.45 \pm 0.45$ | F1280W_hnm |

*Note.* (1) ALMA 1 mm source ID (1mm.C∗: M. Aravena et al. 2020; J. González-López et al. 2020; 3mm.∗: L. A. Boogaard et al. 2019; J. González-López et al. 2019); (2) stellar mass; (3) SFR; (4) dust luminosity; (5) dust mass; (6) configuration from which the physical properties were derived for each source, as described in Section 4.2.2. A machine-readable version of this table, along with the results from all model-restricted configurations (F444W_hn, F1280W_hnm, and F1280W_hnma), is available in the online supplementary material. [a] Sources that are not robustly resolved in both F444W and F1280W are listed with the F444W_hn measurements for completeness.

Comparisons between these dust-based gas masses and CO-based measurements for galaxies with CO line detections (M. Aravena et al. 2019; L. A. Boogaard et al. 2020; D. A. Riechers et al. 2020) are presented in Appendix C. The sample median molecular gas mass is $\log(M_{\mathrm{mol}}/\mathrm{M}_\odot) = 10.26 \pm 0.05$, with the uncertainty estimated via bootstrap resampling.

### *4.2.2 Spatially resolved*

Among the configurations described in Sections 3.2.2 and 3.2.3, we present in Table 3 the optimal integrated physical properties, obtained by summing the measurements across all resolved bins, derived from the model-restricted approach. For each source, the 'optimal' configuration is defined as the one that utilizes the maximum number of filters while maintaining a spatially resolved condition (i.e. the half-light radius in a given filter exceeds twice the PSF half-light radius, as categorized in Table 1). Specifically, the integrated properties for 1mm.C02, 1mm.C03, and 1mm.C10, which are robustly resolved and possess high-resolution ALMA data, are derived from the F1280W_hnma configuration. For other sources resolved at the F1280W resolution, values are taken from F1280W_hnm, while the remaining sources not resolved in F1280W are based on the F444W_hn configuration.

The spatially resolved SED modelling allows us to probe the physical properties across different regions of the galaxies, thereby gaining insights into the star formation physics on kiloparsec (kpc) scales. Fig. 8 presents an example galaxy in the sample, showing the F444W-resolution three-colour image, alongside the physical property maps derived from our F444W_hn SED modelling with the MAGPHYS models restricted as described in Section 3.2.3. The derived physical property maps have a spatial bin size of 0.16 arcsec × 0.16 arcsec per SED modelling element, corresponding to a physical scale of approximately 1.5 kpc for this galaxy. The best-fitting SED of each spatial bin is shown as the thin green curves in Fig. 5.

At first glance, we find that the distributions of $M_*$ and SFR broadly follow the observed light distributions seen in the RGB cut-out. This is consistent with recent spatially resolved SED studies of various galaxy populations using different data sets and modelling codes (Abdurro'uf et al. 2023; C. Giménez-Arteaga et al. 2023; I. Smail et al. 2023; J. Song et al. 2023; J. Li et al. 2024; C. Accard et al. 2025; N. E. P. Lines et al. 2025; Z. Liu et al. 2025; V. Y. Y. Tan et al. 2025; E. Parlanti et al. 2026). Upon closer inspection, the derived physical property maps also reproduce several features visible in the observed light distributions. For example, the red central region of 1mm.C10 corresponds to an older stellar population and higher $A_V$, as shown in the corresponding maps in Fig. 8. On the other hand, the blue clumps in the RGB image correspond to regions of younger stellar populations in the maps. These results illustrate how spatially resolved SED modelling, enabled by high-resolution multiwavelength observations from *HST* and *JWST*, can link observed light distributions to the underlying physical properties.

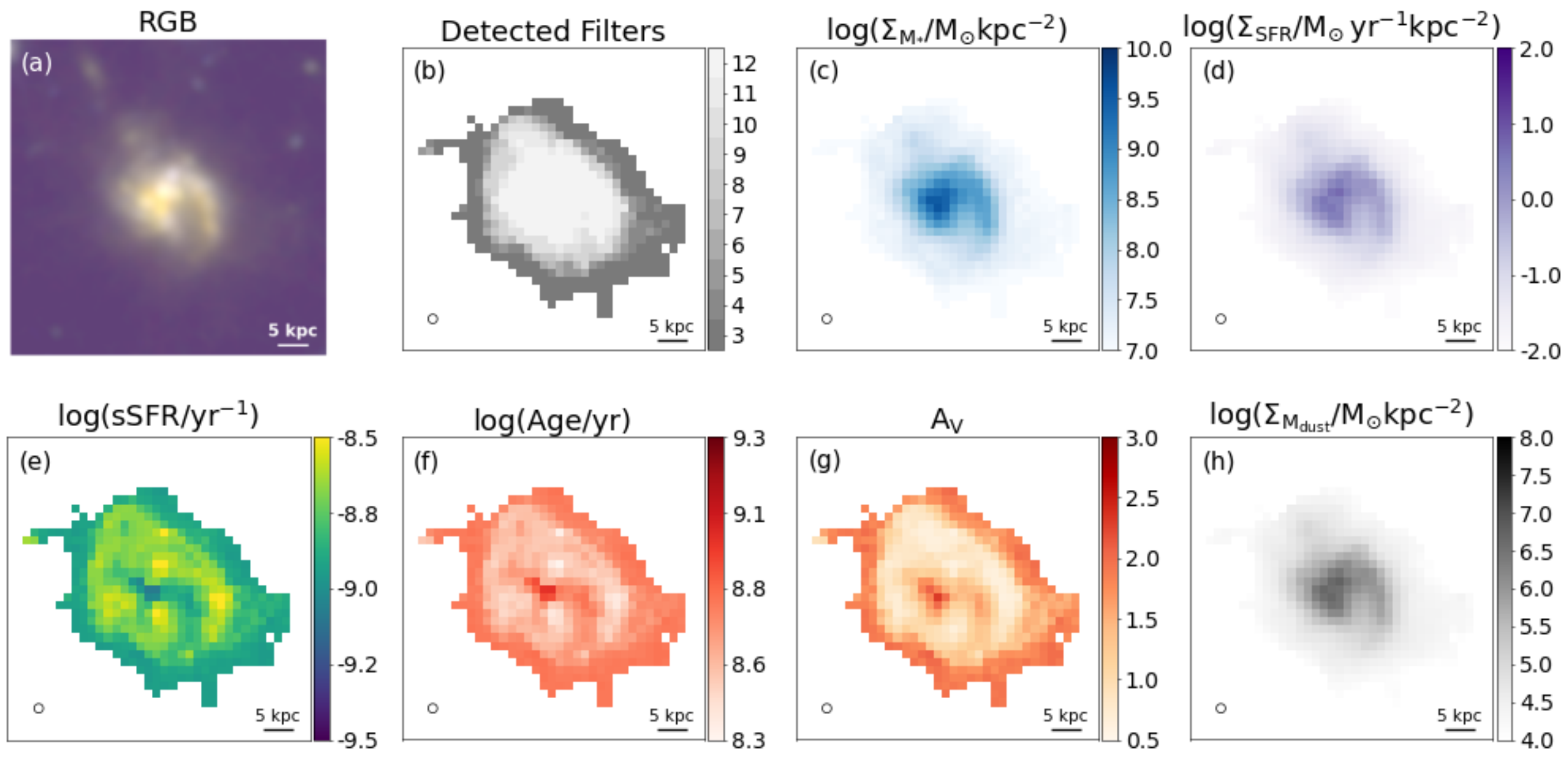

**Figure 8.** Physical property maps of the example galaxy 1mm.C10, derived from the restricted-model F444W_hn SED modelling, corresponding to the green curves presented in Fig. 5. Each map covers a 6 arcsec × 6 arcsec region. Except for the top-left RGB cut-out, all maps have a spatial bin size of 0.16 arcsec × 0.16 arcsec per SED modelling element. The angular resolution is indicated in the bottom-left corner of each panel, and a 5 kpc scale bar is shown in the bottom right. (a) F444W-resolution RGB cut-out displayed using the logarithm stretch. (b) Number of filters that have detections. The remaining panels, from (c) to (h), show the maps of stellar mass surface density ($\Sigma_{M_*}$), SFR surface density ($\Sigma_{SFR}$), specific star formation rate (sSFR), mass-weighted stellar age (age), $V$-band attenuation ($A_V$), and dust mass surface density ($\Sigma_{M_{dust}}$). The corresponding uncertainty maps are presented in Fig. B1, and maps for the full sample are available in the online supplementary material.

In addition to the features discussed above, we also observe patterns in the physical property maps, particularly in stellar age, $A_V$, and sSFR, that are not apparent in the RGB cut-outs. Specifically, these maps show a flat distribution in the outskirts of the galaxies. We note that these areas correspond to the fainter outskirts of the galaxies, where the number of filters with significant detections is lower, as shown in the detected filters map of Fig. 8 (b). Consequently, the derived physical properties in these regions are less constrained, which is reflected in the corresponding uncertainty maps in Fig. B1. This pattern is less evident in the $\Sigma_{M_*}$, $\Sigma_{SFR}$, and $\Sigma_{M_{dust}}$ maps because these quantities are primarily constrained by the total observed flux. As a result, they are less sensitive to the colours, which can be poorly constrained and therefore more strongly influenced by the priors when only a limited number of filters have significant detections.

We note that the MAGPHYS modelling assumes an energy balance between the absorbed UV–optical emission and the re-emitted infrared dust luminosity. While this assumption is generally valid on global or kpc scales, it may be less accurate on smaller spatial scales, where local geometrical effects (e.g. offsets between young stars and dust) can lead to deviations from energy balance. Given the kpc-scale resolution of our analysis, this effect is not expected to dominate our results, but may affect the interpretation of small-scale variations in the spatially resolved properties.

### 4.2.3 *Comparisons of different configurations*

Having derived physical properties from both spatially unresolved and spatially resolved SED modelling, we now compare the results to benchmark our spatially resolved SED modelling pipeline. Our goal is to assess how the inferred physical properties vary with different analysis choices, including (1) spatial resolution (spatially unresolved versus spatially resolved), (2) wavelength coverage (F444W_hn, F1280W_hnm, and F1280W_hnma), and (3) MAGPHYS model libraries (default versus restricted models). These comparisons are designed to test the robustness of the resolved modelling against data handling choices, rather than to explore uncertainties arising from different underlying model assumptions, e.g. IMF (N. Bastian, K. R. Covey & M. R. Meyer 2010; A. M. Hopkins 2018), SFH (A. C. Carnall et al. 2019; J. Leja et al. 2019; S. Lower et al. 2020; R. K. Cochrane et al. 2025), dust attenuation law (S. Salim & D. Narayanan 2020), dust emissivity reference ($\kappa_0$; L. Dunne et al. 2000; B. T. Draine et al. 2014). Among all configurations, we adopt the spatially unresolved modelling as the reference for comparison, as it benefits from the broadest wavelength coverage across the spectrum.

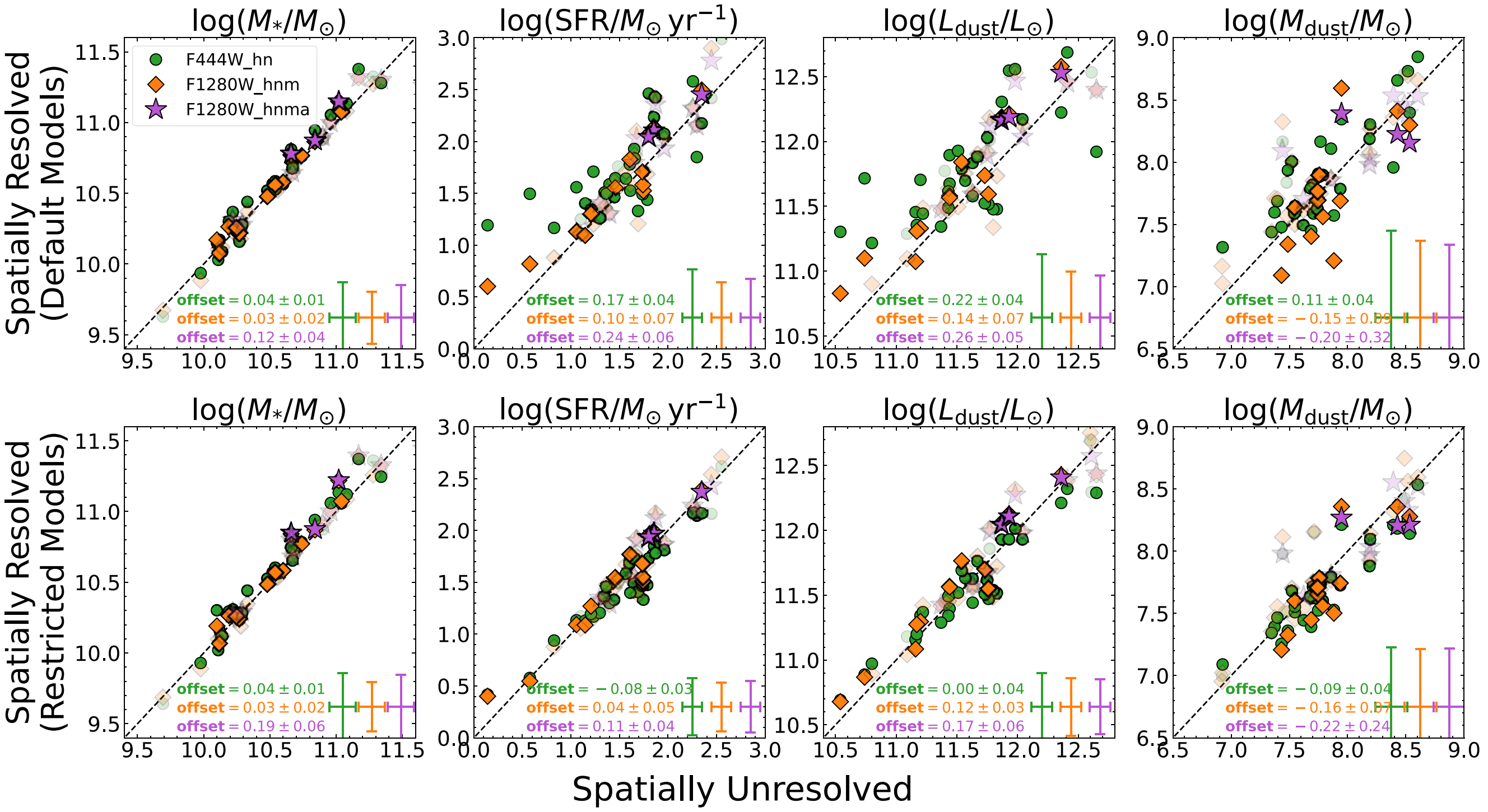

**Figure 9.** Comparison of the MAGPHYS-derived physical properties obtained from different fitting configurations. In all panels, the abscissa shows measurements from the spatially unresolved fitting, while the ordinate shows the integrated values from the spatially resolved modelling, computed as the sum of all spatial bins in the corresponding physical property maps. Different symbols and colours indicate the data coverage used in the spatially resolved fittings (F444W_hn in green circles, F1280W_hnm in orange diamonds, and F1280W_hnma in purple stars). Sources that are not robustly spatially resolved, based on the criteria described in Section 4.3, are plotted transparently. The first and second rows correspond to fittings performed with the default and the restricted MAGPHYS models, respectively. From left to right, the columns show comparisons of $M_*$, SFR, $L_{\rm dust}$, and $M_{\rm dust}$. In each panel, the median offset and the bootstrap uncertainty, for the robustly spatially resolved sources, are indicated at the bottom, and their typical uncertainties are shown in the bottom-right corner. The dashed black line corresponds to the 1:1 relation. As discussed in Section 4.2.3, these comparisons highlight the robustness of the derived $M_*$, the crucial role of MIRI and high-resolution ALMA data in constraining physical properties, and the effectiveness of the model restriction when high-resolution IR data are limited.

Fig. 9 presents a comparison of $M_*$, SFR, $L_{\rm dust}$, and $M_{\rm dust}$ derived from different configurations, and a summary of the main findings from Fig. 9 is as follows:

(i) Spatial resolution: we find that $M_*$ derived from the resolved and unresolved modelling approaches are consistent within $\sim 0.05$ dex. This suggests that our sample does not exhibit a significant $M_*$ discrepancy (see further discussion in Section 5.4). In contrast, the IR-sensitive properties (SFR, $L_{\rm dust}$, and $M_{\rm dust}$) derived using the default models are generally overestimated in the spatially resolved modelling compared to the spatially unresolved case, with the averaged offsets of $\sim 0.17$ dex for SFR, $\sim 0.21$ dex for $L_{\rm dust}$, and $\sim -0.08$ dex for $M_{\rm dust}$ (see Fig. 9 for details on the median offsets across different properties and configurations). However, when the restricted models are applied, the results from spatially resolved and unresolved modelling become consistent within $\sim 0.1$ dex.

(ii) Data coverage: comparing the results from the three spatially resolved configurations (F444W_hn, F1280W_hnm, and F1280W_hnma), we find that including MIRI data is crucial for achieving better consistency with the spatially unresolved results for IR-sensitive properties, especially in the default-model configurations. For example, under the framework of default modelling, adding MIRI photometry reduces the median offset between resolved and unresolved measurements of $L_{\rm dust}$ from 0.22 dex in F444W_hn to 0.14 dex in F1280W_hnm. The inclusion of ALMA data further improves the agreement, with individual F1280W_hnma measurements lying closer to the spatially unresolved values, although the statistical significance is limited by the small sample size (only three sources are spatially resolved). Additional discussions on the impact of MIRI and ALMA data are provided in Sections 5.1 and 5.2, respectively.

(iii) MAGPHYS models: we find that the scatter among data points is smaller in the restricted-model fits, and the results from spatially resolved and unresolved analyses show better agreement. These findings confirm that the model restriction process helps derive consistent physical properties, particularly for IR-sensitive parameters. We further discuss the effects of the model restriction in Section 5.3.

Given that $M_*$ is the most robust quantity, being least affected by the choice of spatial resolution, wavelength coverage, or model usage, our subsequent analyses will primarily focus on $M_*$. Specifically, we will make use of the $\Sigma_{M_*}$ maps derived from the restricted-model F444W_hn configuration, as the derived $M_*$ values do not vary significantly across configurations and F444W_hn offers the highest spatial resolution among all configurtions.

### 4.3 Curve of growth

We perform a curve-of-growth analysis to measure the sizes of $M_*$ and the light profiles across different filters, following approaches commonly adopted in the literature (e.g. C.-C. Chen et al. 2022; J. A. Hodge et al. 2025; J. Chen et al. 2026). This non-parametric method provides model-independent estimates of galaxy sizes, as it does not require assuming a specific functional form for the underlying mass or light distributions. The analysis is carried out using the `CurveOfGrowth` function from the `photutils.profiles` package (M. B. Bagley et al. 2024), adopting concentric elliptical apertures and normalizing to the total flux, which is the flux enclosed within the segmentation map. The aperture centres are fixed to the *JWST* source positions provided by the DJA and JADES catalogues (see Table 1). To determine the ellipticity and PA, we model the F444W image using the `photutils.isophote` package and adopt the parameters of the best-fitting ellipse corresponding to 10 per cent of the peak intensity.

We then derive the effective radii for the stellar mass ($R_{\mathrm{eff,M_*}}$), the light profiles ($R_{\mathrm{eff,filter}}$), and the PSF ($R_{\mathrm{PSF,filter}}$), defined as the semi-major axis at which the curve-of-growth profile reaches 50 per cent of the total mass or flux. We note that the effective radii derived from the curve-of-growth profiles are not corrected for the PSF effects. For the spatially resolved sources considered in this study, the influence of the PSF on the measured sizes is expected to be small ($\lesssim$ 10 per cent). Moreover, applying a PSF correction would require assuming a parametric form for the underlying light or stellar mass distribution, which would not be aligned with the non-parametric nature of our measurements.

To assess whether a source is robustly spatially resolved, we require the measured $R_{\mathrm{eff,filter}}$ to be at least twice the effective radius of the PSF in the corresponding filter ($R_{\mathrm{PSF,filter}}$). The resolved or unresolved classification for each source at the F444W and F1280W filters is summarized in Table 1. Based on this criterion, we restrict the subsequent analysis to sources that are robustly spatially resolved. Out of our 35 galaxies, 31 are resolved in F444W and 14 in F1280W.

The derived effective radii of $M_*$ are listed in Table 4, and their distributions from the model-restricted F444W_hn and F1280W_hnm modelling, together with those measured from different *JWST* filters, are shown in Fig. 10. We note that although the galaxy pair 1mm.C23 is not deblended, we also measure and report its size for completeness. However, this source is excluded from the size-related scaling-relation analyses in the subsequent sections, as explicitly noted in the corresponding figure captions. In both panels of Fig. 10, only robustly resolved sources are shown, which explains the observed size offset between the two panels as the sizes in the F1280W_hnm plot are biased to the larger sources. We find that the effective radii of the light profiles decreases with increasing wavelength, consistent with previous studies of observed and simulated galaxies from local up to $z \sim 5$ (V. Casasola et al. 2017; C.-C. Chen et al. 2022; G. Popping et al. 2022; K. A. Suess et al. 2022; R. K. Cochrane et al. 2023; M. Baes et al. 2024; S. Gillman et al. 2024; J. A. Hodge et al. 2025). This trend likely reflects a combination of intrinsic wavelength-dependent galaxy structure and the effects of dust obscuration. Moreover, we find that effective radii measured at the rest-frame $\sim$ 1–3 µm (e.g. F444W and F560W) generally provide a good proxy for $R_{\mathrm{eff,M_*}}$, supporting the commonly adopted assumption that rest-frame NIR light effectively traces stellar mass (A. Saintonge et al. 2017; J. Freundlich et al. 2019; U. Lisenfeld et al. 2023; L. A. Boogaard et al. 2024).

**Table 4.** Source properties.

| ID (1) | $R_{\mathrm{eff,M_*}}$/kpc (2) | $\log_{10}(\Sigma_{\mathrm{eff},*}/\mathrm{M_\odot\ kpc^{-2}})$ (3) |
|---|---|---|
| 1mm.C01 | $3.5 \pm 0.2$ | $8.4 \pm 0.2$ |
| 1mm.C02 | $5.0 \pm 0.3$ | $8.6 \pm 0.4$ |
| 1mm.C03 | $4.8 \pm 0.3$ | $9.0 \pm 0.1$ |
| 1mm.C04 | $2.6 \pm 0.2$ | $9.1 \pm 0.1$ |
| 1mm.C05 | $2.9 \pm 0.3$ | $9.7 \pm 0.1$ |
| 1mm.C06 | $2.2 \pm 0.2$ | $9.8 \pm 0.1$ |
| 1mm.C07 | $1.6 \pm 0.2$ | $9.7 \pm 0.2$ |
| 1mm.C08 | $1.0 \pm 0.1$ | $10.5 \pm 0.2$ |
| 1mm.C09 | $2.4 \pm 0.2$ | $8.5 \pm 0.2$ |
| 1mm.C10 | $4.3 \pm 0.2$ | $8.9 \pm 0.2$ |
| 1mm.C11 | $1.8 \pm 0.3$ | $9.3 \pm 0.2$ |
| 1mm.C12 | $5.7 \pm 0.5$ | $8.0 \pm 0.4$ |
| 1mm.C13 | $2.1 \pm 0.2$ | $9.6 \pm 0.1$ |
| 1mm.C14a | $2.0 \pm 0.2$ | $9.1 \pm 0.2$ |
| 1mm.C14b | $4.2 \pm 0.3$ | $8.3 \pm 0.3$ |
| 1mm.C15 | $2.7 \pm 0.3$ | $9.4 \pm 0.2$ |
| 1mm.C16 | $4.2 \pm 0.2$ | $8.5 \pm 0.4$ |
| 1mm.C17 | $2.7 \pm 0.3$ | $8.9 \pm 0.3$ |
| 1mm.C18 | $4.4 \pm 0.3$ | $8.4 \pm 0.4$ |
| 1mm.C19 | $3.1 \pm 0.2$ | $8.8 \pm 0.2$ |
| 1mm.C20 | $3.5 \pm 0.3$ | $8.9 \pm 0.3$ |
| 1mm.C21 | $3.1 \pm 0.3$ | $8.5 \pm 0.3$ |
| 1mm.C22 | $3.8 \pm 0.3$ | $8.3 \pm 0.4$ |
| 1mm.C23[a] | $7.5 \pm 0.2$ | $8.2 \pm 0.5$ |
| 1mm.C24 | $3.9 \pm 0.3$ | $8.2 \pm 0.2$ |
| 1mm.C25 | $2.0 \pm 0.2$ | $9.2 \pm 0.1$ |
| 1mm.C26 | $2.8 \pm 0.2$ | $8.6 \pm 0.1$ |
| 1mm.C28 | $6.6 \pm 0.2$ | $8.1 \pm 0.5$ |
| 1mm.C30 | $1.9 \pm 0.1$ | $8.3 \pm 0.4$ |
| 1mm.C31 | $4.2 \pm 0.2$ | $7.9 \pm 0.4$ |
| 1mm.C32 | $4.2 \pm 0.3$ | $8.2 \pm 0.3$ |
| 1mm.C33 | $4.3 \pm 0.4$ | $9.0 \pm 0.4$ |
| 3mm.09 | $1.3 \pm 0.1$ | $10.2 \pm 0.1$ |
| 3mm.11 | $3.3 \pm 0.2$ | $8.3 \pm 0.4$ |
| 3mm.16 | $3.6 \pm 0.3$ | $8.4 \pm 0.4$ |

*Notes.*
(1) ALMA 1 mm source ID (1mm.C∗: M. Aravena et al. 2020; J. González-López et al. 2020; 3mm.∗: L. A. Boogaard et al. 2019; J. González-López et al. 2019)
(2) Effective stellar mass radii ($R_{\mathrm{eff},*}$) measured from the model-restricted F444W_hn modelling (Section 4.3)
(3) Effective stellar mass surface density ($\Sigma_{\mathrm{eff},*}$), as described in Section 4.3
[a] For completeness, we also report the measurements of the pair galaxy 1mm.C23 here.

### 4.4 Stellar mass surface density

Based on the stellar mass sizes derived from the curve-of-growth analysis described above, we compute the effective stellar mass surface density ($\Sigma_{\mathrm{eff},*}$) as

$$\Sigma_{\mathrm{eff},*} = \frac{M_*}{2\pi R^2_{\mathrm{eff,M_*}}}, \tag{2}$$

where half of the total stellar mass is divided by the area enclosed within the half-mass radius. Both $M_*$ and $R_{\mathrm{eff,M_*}}$ are adopted from the restricted-model F444W_hn modelling. The estimated $\Sigma_{\mathrm{eff},*}$ for each galaxy is listed in Table 4.

The effective stellar mass surface density serves as a quantitative indicator of galaxy morphology, with higher $\Sigma_{\mathrm{eff},*}$ generally reflecting a more bulge-dominated system (A. Saintonge & B. Catinella 2022; U. Lisenfeld et al. 2023). P. F. Hopkins et al.

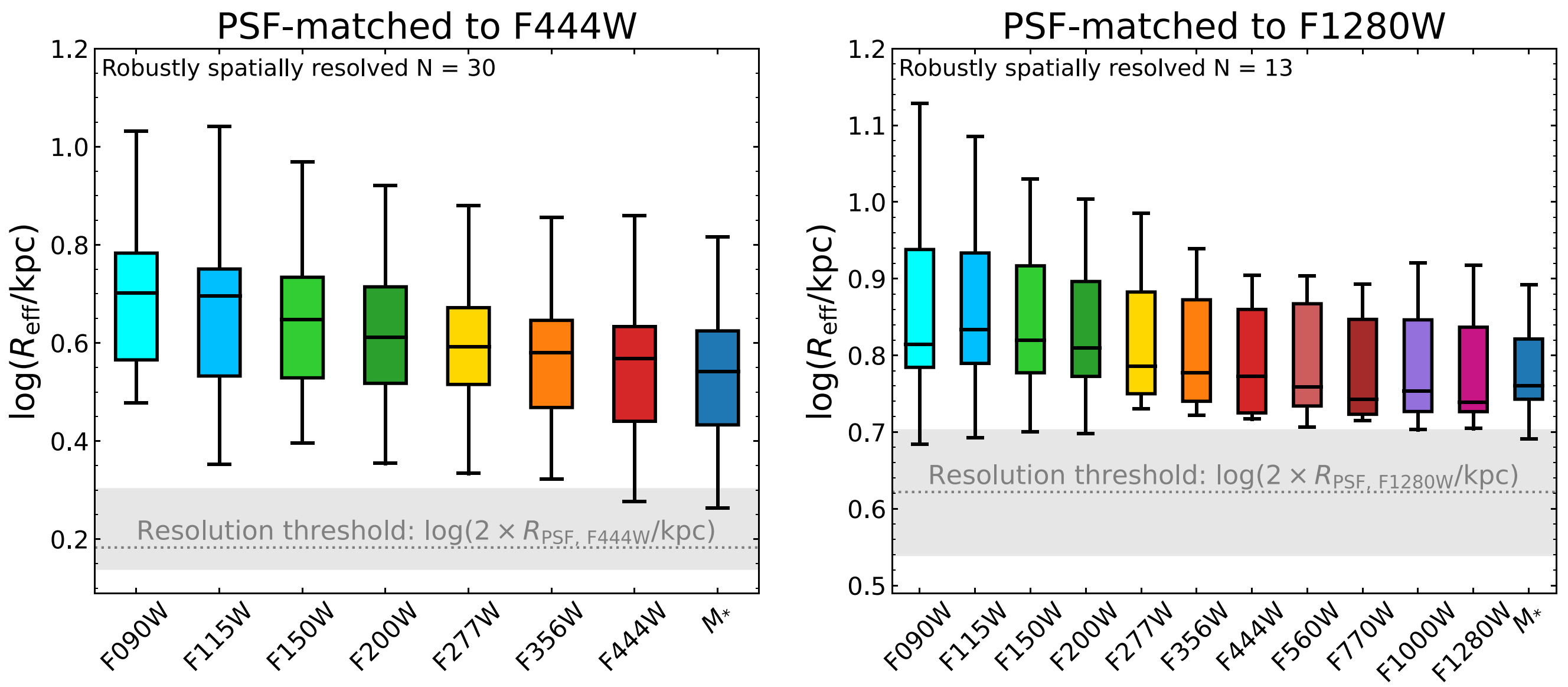

**Figure 10.** Box plots of effective radii ($R_{\rm eff}$) measured from the $M_*$ distribution and from light profiles in different *JWST* filters. The left panel shows results from the F444W_hn modelling, and the right panel presents those from the F1280W_hnm modelling. Only sources that are robustly spatially resolved are included, defined as having sizes larger than twice the PSF size (see Section 4.3). 1mm.C23 is excluded from the plot due to the spatial overlap of the galaxy pair. The number of sources in each case is indicated in the upper-left corner. The grey dotted line shows the resolution threshold at the median redshift, with the shaded region indicating the range spanned by the sample. In each box plot, the box spans the first and third quartiles, with the median indicated by a horizontal black line, and the whiskers extend to data points within 1.5 times the interquartile range (IQR). The stellar mass effective radii are generally consistent with those measured from the F444W and F560W light, which trace rest-frame $\sim 1.5$–$2.0$ μm for galaxies at cosmic noon.

(2010) suggest that the maximum $\Sigma_{\rm eff,*}$ in galaxies is limited to $\sim 10^{11}\ {\rm M_\odot\ kpc^{-2}}$, regulated by feedback from massive stars. We find that our measurements, both for individual spatially resolved bins and the effective values for each galaxy, are lower than this limit.

Furthermore, $\Sigma_{\rm eff,*}$ provides a crucial link between the stellar distribution and the gas properties in galaxies, as its relationship with molecular gas offers insights into star formation mechanisms. Through the resolved SED modelling, we are able to robustly measure $\Sigma_{\rm eff,*}$, which serves as an additional parameter for understanding the physical processes that govern galaxy formation and evolution.

## 5 DISCUSSION

### 5.1 The impact of including MIRI data

Most previous spatially resolved SED modelling studies of non-local samples have relied primarily on *HST* and NIRCam data, which probe mainly the rest-frame UV-to-optical regime. Only a few works incorporate longer wavelength observations, such as the study of two lensing submillimetre sources (I. Smail et al. 2023), which uses the 880 μm maps from the Submillimetre Array, the study of the ALMA-CRISTAL sample (J. Li et al. 2024), which includes ALMA submillimetre data, and the ALESS submillimetre galaxy study (J. Li et al. 2026), which combines MIRI F770W and ALMA 870 μm dust continuum. In this context, our spatially resolved SED modelling, utilizing MIRI observations up to F1280W, provides a valuable data set to assess the impact of MIRI on constraining galaxy physical properties.

In our default-model configurations, we find that including MIRI data in the spatially resolved SED modelling is crucial for accurately constraining IR-sensitive physical parameters, such as $L_{\rm IR}$, SFR, and $M_{\rm dust}$. As shown in the upper panels of Fig. 9, incorporating rest-frame NIR data from MIRI significantly reduces the overestimation of these IR-sensitive properties compared to the values from the spatially unresolved fitting, roughly by 20 per cent on average, and in some cases, the reduction can be up to $\sim 0.8$ dex. While a modest overestimation of $\sim 0.15$ dex remains, possibly due to the limited constraints on the dust emission peak in the far-IR ($\sim$ a few hundred μm), the MIRI-included estimates are considerably more accurate than those obtained using only *HST* and NIRCam data. Our findings are consistent with the extensive spatially unresolved SED fitting tests of C. Pacifici et al. (2023), who report a systematic overestimation of SFR when the rest-frame IR data are excluded, as well as with U. Dudzevičiūtė et al. (2020), who also employ MAGPHYS for SED modelling. This highlights the importance of including IR observations when deriving IR-sensitive physical properties, and suggests caution when interpreting SED results lacking such coverage.

For the IR-sensitive properties derived from the restricted-model SED fittings (lower panels in Fig. 9), we find that the inclusion of MIRI observations has a minimal or an opposite impact. This is not surprising, as the restricted-model dust emission models are already reasonably constrained by the global colours, so the MIRI data only provide little additional information. Further discussion and caveats regarding the effects of model restriction are presented in Section 5.3.

We find that the impact of MIRI data on $M_*$ is less pronounced than on the IR-sensitive properties, for both default- and restricted-model configurations. This suggests that NIRCam-only modelling is generally sufficient to estimate $M_*$ for the typical main-sequence galaxies at $z \sim 2$, although including MIRI observations still leads to slightly improved measurements. For

higher-$z$ samples, the inclusion of the MIRI data becomes more important, as the rest-frame NIR light shifts into its wavelength coverage. A more detailed discussion of $M_*$ comparisons, along with a summary of findings from recent spatially resolved SED fitting studies, is provided in Section 5.4.

One might argue that the effects discussed above could be influenced by both wavelength coverage and resolution, since including the MIRI data not only extends the wavelength coverage but also changes the effective angular resolution. To test this, we re-run the spatially resolved fittings using only *HST* and NIRCam data, but PSF-matched to the MIRI F1280W resolution (namely the F1280W_hn modelling). We find that the F1280W_hn modelling results are consistent with those from the F444W_hn modelling, indicating that the impact of MIRI arises primarily from the broader wavelength coverage rather than the change of angular resolution.

We also perform the same SED modelling using another widely adopted code, CIGALE[5] (M. Boquien et al. 2019), which implements energy balance in a manner similar to MAGPHYS. We constructed the CIGALE priors to closely match the MAGPHYS parameters and find good consistency between the physical properties derived from the two codes, with the only notable exception being $M_{\rm dust}$, which reflects differences in the assumed dust emissivity reference ($\kappa_0$), as discussed in C.-L. Liao et al. (2024). This indicates that CIGALE, similar to MAGPHYS, tends to overpredict the FIR emission when adopting the same priors. Overall, these tests confirm that our conclusions regarding the impact of MIRI observations remain robust.

### 5.2 The impact of including ALMA data

Some galaxies in our sample have one or two submillimetre data from ALMA with a resolution comparable to that of F1280W (Section 2.3). Including these data allows us to better constrain the Rayleigh–Jeans tail of the dust emission and to directly compare SED modelling results with and without ALMA constraints. Fig. 11 illustrates how the derived physical properties differ between the F1280W_hnma and F1280W_hnm modelling as a function of radius. With the exception of $M_*$, which shows minimal sensitivity to the inclusion of ALMA data, other properties exhibit more pronounced differences toward the central regions, with larger offsets seen in the default-model fitting than in the restricted-model case. These results indicate that the inclusion of ALMA data can lead to changes of up to $\sim 0.1$ dex in IR-sensitive properties in the central regions of main-sequence galaxies at cosmic noon.

Consistent with findings from UV-selected galaxies at higher redshift (J. Li et al. 2024), our results also show that incorporating ALMA observations systematically lowers the inferred $A_{\rm V}$, while increasing the mass-weighted stellar age. This highlights the crucial role of ALMA data in mitigating the age–dust degeneracy: the observed red colours can be more reliably attributed to older stellar populations rather than to dust extinction.

We further find that incorporating high-resolution ALMA data without MIRI (i.e. the F1280W_hna configuration) yields a similar overall reduction in the overestimation of IR-sensitive properties as observed in the F1280W_hnm (MIRI-only) set-up. While both F1280W_hnm and F1280W_hna achieve comparable reduction to the integrated $L_{\rm dust}$, MIRI and ALMA constrain fundamentally distinct regimes of the dust SED. MIRI primarily targets the rest-frame NIR/MIR regime, constraining warm dust emission and PAH features, whereas ALMA anchors the FIR Rayleigh–Jeans tail, which governs the cold dust mass. As a consequence, omitting either data set alters the derived SED shape, demonstrating that neither MIR nor FIR constraints alone are sufficient to fully characterize the IR SED and correctly infer the properties. These findings underline the distinct and highly complementary roles of MIRI and ALMA in constraining dust-related physical properties.

We note that the results discussed here are based on the subset of galaxies for which matched-resolution ALMA data are available. Extending this analysis to the full ASPECS sample will require additional high-resolution ALMA observations to test whether the trends identified here persist more generally. Nevertheless, our results highlight that FIR constraints are crucial for disentangling stellar age and dust attenuation, and for obtaining robust dust-related physical properties. Given that high-resolution FIR data are not available for all galaxies in the sample, we focus our physical interpretation in the remainder of the paper on $M_*$-related quantities only.

### 5.3 Effectiveness and caveat of MAGPHYS model restriction

Restricting the MAGPHYS model library using spatially unresolved colours (Section 3.2.3) provides an effective approach to constrain the dust SED in the absence of high-resolution MIR and FIR data. As demonstrated by the agreement ($\lesssim 0.1$ dex) between the summed resolved $L_{\rm dust}$ and unresolved measurements, this method successfully confines the resolved modelling to match the unresolved modelling. The effectiveness is particularly valuable for configurations lacking MIRI coverage (e.g. F444W_hn).

Importantly, the application of model restriction does not significantly degrade the quality of the SED modelling. We find that the results obtained using the restricted models achieve comparable goodness-of-fit to those derived from the default model library, as quantified by similar reduced $\chi^2$ values across spatial bins. This indicates that restricting the model space using observationally motivated colour constraints does not force poorer fits to the data, but instead removes models that are inconsistent with the global FIR properties while retaining those that adequately reproduce the observed SEDs. In this sense, the default-model configuration provides a physically reasonable baseline, while the restricted-model configuration represents a refinement that incorporates additional empirical constraints without compromising fit quality.

Despite its effectiveness, some care is needed when applying the model-restricting procedure. The additional colour constraints, derived from spatially unresolved photometry, may limit the flexibility of the spatially resolved modelling by encouraging the SED shapes of the resolved models to follow that of the unresolved measurements within an acceptance range. Although adopting a narrower colour acceptance range improves the consistency between the summed spatially resolved properties and the unresolved measurements, it also reduces the freedom of the models to capture potential spatial variations in dust temperature and emission. Similarly, the number and choice of unresolved colours used in the restriction can influence the degree of this effect. As a result, although the integrated physical properties across galaxies remain robust, the detailed spatial distributions inferred from the restricted models should be interpreted with appropriate caution.

[5]https://cigale.lam.fr/

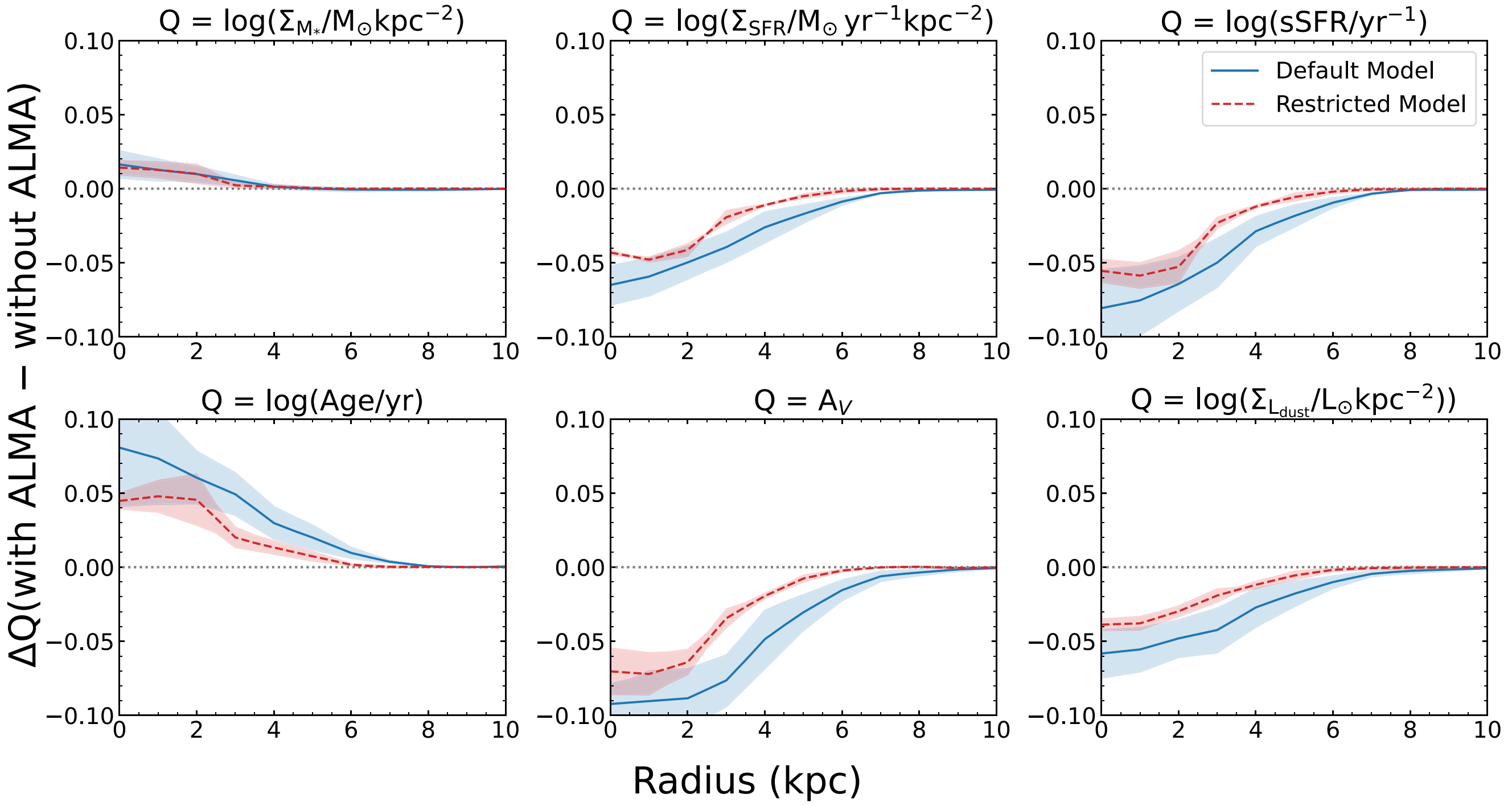

**Figure 11.** Radial profiles of the differences in physical quantities between modelling with ALMA (F1280W_hnma) and without ALMA (F1280W_hnm). The blue solid and red dashed lines show the mean profiles from the default and restricted models, respectively, with shaded regions indicating the 16th–84th percentile ranges. Only the three sources (1mm.C02, 1mm.C03, and 1mm.C10) that are robustly resolved in F1280W and have high-resolution ALMA data are included. The inclusion of ALMA data leads to older stellar ages in the central regions, while $\Sigma_{\rm SFR}$, sSFR, $A_{\rm V}$, and $\Sigma_{\rm L_{dust}}$ are systematically reduced. This effect is more pronounced in the default model fitting and highlights the role of ALMA data in breaking the age–dust degeneracy.

### 5.4 Minimal stellar mass discrepancy in ASPECS galaxies

Recent spatially resolved SED modelling studies (C. Giménez-Arteaga et al. 2023, 2024) reveal that $M_*$ estimates sometimes conflict between the resolved and unresolved estimates in various galaxy samples at $z \gtrsim 3$. The underestimation of the spatially unresolved $M_*$ is estimated to be $\gtrsim 0.5$ dex for the $5 < z < 9$ lensed galaxies in cluster fields (C. Giménez-Arteaga et al. 2023) and for a strongly lensed $z = 6.072$ galaxy (C. Giménez-Arteaga et al. 2024). The spatial resolutions in these studies are all higher than 3 kpc, consistent with the characteristic resolution required to reveal the outshining effect proposed by R. Sorba & M. Sawicki (2015). Moreover, the longest wavelength data used in these studies are all shorter than rest-frame 1 μm, which aligns with the conclusion presented in J. Song et al. (2023). Another possible explanation is that these low-mass ($M_* \lesssim 10^9$ M$_\odot$) galaxies have relatively shallow gravitational potentials compared to more massive systems, making their gas more prone to instability and leading to more recent and ongoing star formation. This results in a larger contribution from young stellar populations to the observed light, thereby amplifying the discrepancy in $M_*$ estimates.

Conversely, some studies (I. Smail et al. 2023; J. Li et al. 2024; L. Shen et al. 2024; N. E. P. Lines et al. 2025) report no significant mass discrepancy between spatially resolved and unresolved SED fittings for galaxies with $M_* \gtrsim 10^9$ M$_\odot$. In J. Li et al. (2024), 14 star-forming main-sequence CRISTAL galaxies at $4 < z < 6$ are modelled using high-resolution data from *JWST* NIRCam filters and ALMA Band 7 [full width at half-maximum (FWHM) $\approx 0.5$ arcsec , corresponding to $\approx 3.2$ kpc at $z = 3$]. The lack of a mass discrepancy may result from the angular resolution being insufficient (e.g. $\lesssim 3$ kpc) to fully separate young and old stellar populations. L. Shen et al. (2024) focus on 19 star-forming galaxies at $0.6 < z < 2.2$ using *HST* and *JWST* NIRCam data for spatially resolved SED fitting. Despite achieving high angular resolution (*HST* F160W FWHM= 0.18 arcsec, $\sim 1.5$ kpc at $z \sim 2$) capable of separating stellar populations, no strong mass discrepancy is observed. This is likely because the reddest filter used in their spatially resolved fittings probes wavelengths longer than rest-frame 1 μm, as proposed by J. Song et al. (2023).

In our study, we do not detect a significant discrepancy of $M_*$ for our cosmic noon galaxies. As shown in Fig. 9, $M_*$ derived from spatially resolved modelling are consistent with those from spatially unresolved modelling, with offsets of $\lesssim 0.1$ dex across all configurations (although the F1280W_hnma case may be slightly affected by the small sample size). We note that the sources with the largest unresolved mass underestimation ($\sim 0.2$ dex) tend to be those with higher $M_{\rm dust}$ and $A_V$, consistent with the findings of J. Li et al. (2026). One possible explanation for the minimal mass discrepancy in our sample is that it does not primarily consist of low-mass galaxies dominated by young stellar populations, where such discrepancies are found to be more prominent in low-mass systems (C. Giménez-Arteaga et al. 2023, 2024; N. E. P. Lines et al. 2025). In addition, they are not strongly dust-obscured, inferring that spatial variations in dust extinction may not strongly deviate from the global properties, thereby limiting the impact of dust attenuation on the inferred $M_*$. Another contributing factor might be that our modelling includes data extending to the rest-frame NIR. As shown in J. Song et al. (2023), $M_*$ estimates become largely unbiased when the reddest filter probes wavelengths longer than rest-frame 1 μm. In our sample, even in the most extreme case, the F444W_hn modelling of the highest-redshift galaxy (1mm.C09 at $z = 3.601$), still roughly holds. Furthermore, including MIRI data in the spatially resolved modelling reduces the median $M_*$ offset by 0.01 dex for both default- and restricted-model fits. This subtle improvement likely reflects the combined effect of additional rest-frame NIR constraints and the coarser spatial resolution, from F444W_hn (0.18 arcsec, $\sim 1.5$ kpc at $z = 2$) to F1280W_hnm (0.50 arcsec, $\sim 4.2$ kpc at $z = 2$).

### 5.5 Relations between molecular gas and $\Sigma_{\mathrm{eff},*}$

We investigate how the effective stellar mass surface density ($\Sigma_{\mathrm{eff},*} = M_*/2\pi R^2_{\mathrm{eff,M_*}}$) relates to the global molecular gas mass ($M_{\mathrm{gas,mol}}$), molecular gas-to-stellar mass ratio ($\mu_{\mathrm{gas,mol}} = M_{\mathrm{gas,mol}}/M_*$), and molecular gas depletion time ($t_{\mathrm{depl,mol}} = M_{\mathrm{gas,mol}}/\mathrm{SFR}$, which is the inverse of the star formation efficiency). The molecular gas mass, $M_{\mathrm{gas,mol}}$, is derived as described in Section 4.2.1, while the $M_*$ and SFR are taken from the spatially unresolved SED modelling. This approach allows us, for the first time, to directly connect internal structural parameters to global gas and star formation properties in typical galaxies at cosmic noon, providing a cosmic noon counterpart to measurements in the local Universe (A. K. Leroy et al. 2008; Y. Shi et al. 2011; A. Boselli et al. 2014; M.-L. Huang & G. Kauffmann 2014; A. Saintonge et al. 2017; B. Catinella et al. 2018; J. Freundlich et al. 2019; V. Villanueva et al. 2021; A. Saintonge & B. Catinella 2022; U. Lisenfeld et al. 2023).

In Fig. 12, we present the correlations between $M_{\mathrm{gas,mol}}$ and $\mu_{\mathrm{gas,mol}}$ as a function of $\Sigma_{\mathrm{eff},*}$ of our $z \sim 1.85$ ASPECS galaxies that are robustly spatially resolved in the F444W resolution. Local measurements from the $z \leq 0.05$ xCOLD GASS sample (A. Saintonge et al. 2017) are also plotted as a comparison. We performed a Spearman rank correlation test, with the correlation coefficient ($r_s$) and the associated $p$-value ($p$) displayed in the bottom-left corner of each panel in Fig. 12. The results show a statistically significant positive $M_{\mathrm{gas,mol}}$–$\Sigma_{\mathrm{eff},*}$ correlation in both samples. Furthermore, $\mu_{\mathrm{gas,mol}}$ is significantly correlated with $\Sigma_{\mathrm{eff},*}$ in the xCOLD GASS sample and marginally correlated in the ASPECS sample, likely owing to the limited sample size. We further perform non-linear least-squares fits to the individual measurements of both samples in logarithmic space, adopting a simple linear relation normalized at $\log_{10}(\Sigma_{\mathrm{eff},*}/\mathrm{M}_\odot\,\mathrm{kpc}^{-2}) = 9$. Median values binned in $\Sigma_{\mathrm{eff},*}$ are also shown, with the points colour-coded by their median offset from the SFMS ($\Delta$MS). Similar comparisons of the correlations between $t_{\mathrm{depl,mol}}$ and $\Sigma_{\mathrm{eff},*}$ for both samples are presented in Fig. 13.

Based on the best-fitting normalization, our $z \sim 1.85$ ASPECS galaxies exhibit values ~1.3 dex higher in $M_{\mathrm{gas,mol}}$, ~1.0 dex higher in $\mu_{\mathrm{gas,mol}}$, and ~0.2 dex lower in $t_{\mathrm{depl,mol}}$, on average, compared to local galaxies. These offsets are consistent with the more gas-rich conditions at cosmic noon, under which galaxies accumulate larger molecular gas reservoirs and form stars more efficiently than their local counterparts, in line with the expected redshift evolution of $\mu_{\mathrm{gas,mol}}$ and $t_{\mathrm{depl,mol}}$ (e.g. L. J. Tacconi et al. 2020). In addition, for both samples, we find a mild decrease in $\mu_{\mathrm{gas,mol}}$ as well as an increase in $M_{\mathrm{gas,mol}}$ and $t_{\mathrm{depl,mol}}$ with increasing $\Sigma_{\mathrm{eff},*}$. These trends suggest that galaxies with higher $\Sigma_{\mathrm{eff},*}$, which are typically located lower on the SFMS (indicated by the colours), have relatively reduced gas-to-stellar mass ratio and longer gas depletion times, consistent with properties associated with more evolved galaxies.

The similar best-fitting slopes of the $M_{\mathrm{gas,mol}}$–$\Sigma_{\mathrm{eff},*}$ and $\mu_{\mathrm{gas,mol}}$–$\Sigma_{\mathrm{eff},*}$ relations in the ASPECS and xCOLD GASS samples indicate that the dependence of $M_{\mathrm{gas,mol}}$ and $\mu_{\mathrm{gas,mol}}$ on $\Sigma_{\mathrm{eff},*}$ is comparable in the two samples. Although galaxies at cosmic noon host systematically larger molecular gas reservoirs than local galaxies at fixed $\Sigma_{\mathrm{eff},*}$, the relative change in $M_{\mathrm{gas,mol}}$ and $\mu_{\mathrm{gas,mol}}$ with $\Sigma_{\mathrm{eff},*}$ are not dramatically different at $z \sim 1.85$ and $z \sim 0$. This implies that the relative role of $\Sigma_{\mathrm{eff},*}$ in regulating the molecular gas reservoir does not significantly change across cosmic time. Similarly, the slopes of the $t_{\mathrm{depl,mol}}$–$\Sigma_{\mathrm{eff},*}$ relations are consistent within the uncertainties. This indicates that there is no clear evidence for a different relative change of $t_{\mathrm{depl,mol}}$ in $\Sigma_{\mathrm{eff},*}$ between galaxies at cosmic noon and those in the present-day Universe.

Beyond the best-fitting linear relations, we identify a subtle deviation at $\log_{10}(\Sigma_{\mathrm{eff},*}/\mathrm{M}_\odot\mathrm{kpc}^{-2}) \gtrsim 9$ in the binned $\mu_{\mathrm{gas,mol}}$–$\Sigma_{\mathrm{eff},*}$ relation for the two samples. In the local galaxies, $\mu_{\mathrm{gas,mol}}$ decreases more steeply in this regime, which A. Saintonge et al. (2017) attribute to the bulge-dominated galaxies. In contrast, the binned ASPECS data exhibit a flatter trend at similar $\Sigma_{\mathrm{eff},*}$. This difference may indicate a suppression of star formation at higher stellar surface densities without a rapid depletion of the molecular gas reservoir, which would imply morphological quenching plays a more prominent role in more evolved galaxies at cosmic noon. Alternatively, high-$\Sigma_{\mathrm{eff},*}$ galaxies at cosmic noon may remain gas-rich despite being quenched, as continued gas accretion from their gas-rich environments could sustain their gas reservoirs. In addition, we identify a tentative turnover in the $t_{\mathrm{depl,mol}}$–$\Sigma_{\mathrm{eff},*}$ trend for galaxies above and below the SFMS in both samples. The physical origin of this turnover remains unclear, and larger samples will be required to robustly constrain this behaviour.

Lastly, in Fig. 13, we include the $t_{\mathrm{depl,mol}}$–$\Sigma_{\mathrm{eff},*}$ relations derived from galaxies with $M_* > 10^9\,\mathrm{M}_\odot$ and SFR $> 0$ in the COLIBRE hydrodynamic simulation (C. d. P. Lagos et al. 2026) at $z = 0$ and $z = 2$. The overall similarity in the shape of the simulated relations at both redshifts suggests that the impact of $\Sigma_{\mathrm{eff},*}$ on the relative variation of the molecular gas depletion time does not evolve strongly with cosmic time. This is consistent with the trends inferred from the ASPECS and xCOLD GASS samples. When comparing the observations with the simulations, we do not find evidence for the flattening feature seen in the simulated relations. This discrepancy may arise from observational incompleteness and selection bias, particularly the lack of compact galaxies (e.g. $R_e < 2$ kpc) in our sample. In the low-$\Sigma_{\mathrm{eff},*}$ regime, the agreement between observations and simulations is better, as galaxies in this regime are typically more extended. In contrast, the high-$\Sigma_{\mathrm{eff},*}$ regime is expected to be dominated by more compact galaxies that display disordered motions, which may be underrepresented in the observations that required sufficient resolution to drive result properties. These compact galaxies tend to have relatively shallow stellar gravitational potentials, making their gas more prone to collapse and leading to shorter $t_{\mathrm{depl,mol}}$ (E. Wang, X. Kong & Z. Pan 2018; J. Chu et al. 2026; Z. He et al. 2026). As a result, the bias toward larger galaxies in the observed sample, due to the requirement to derive resolved properties, may explain why the flattening or declining trend predicted by the simulations is not observed.

Our results provide, for the first time, a qualitative examination of the role of $\Sigma_{\mathrm{eff}*}$ in regulating the molecular gas reservoir, molecular gas-to-stellar mass ratio, and star formation efficiency beyond the local Universe using a uniformly analyzed data set.

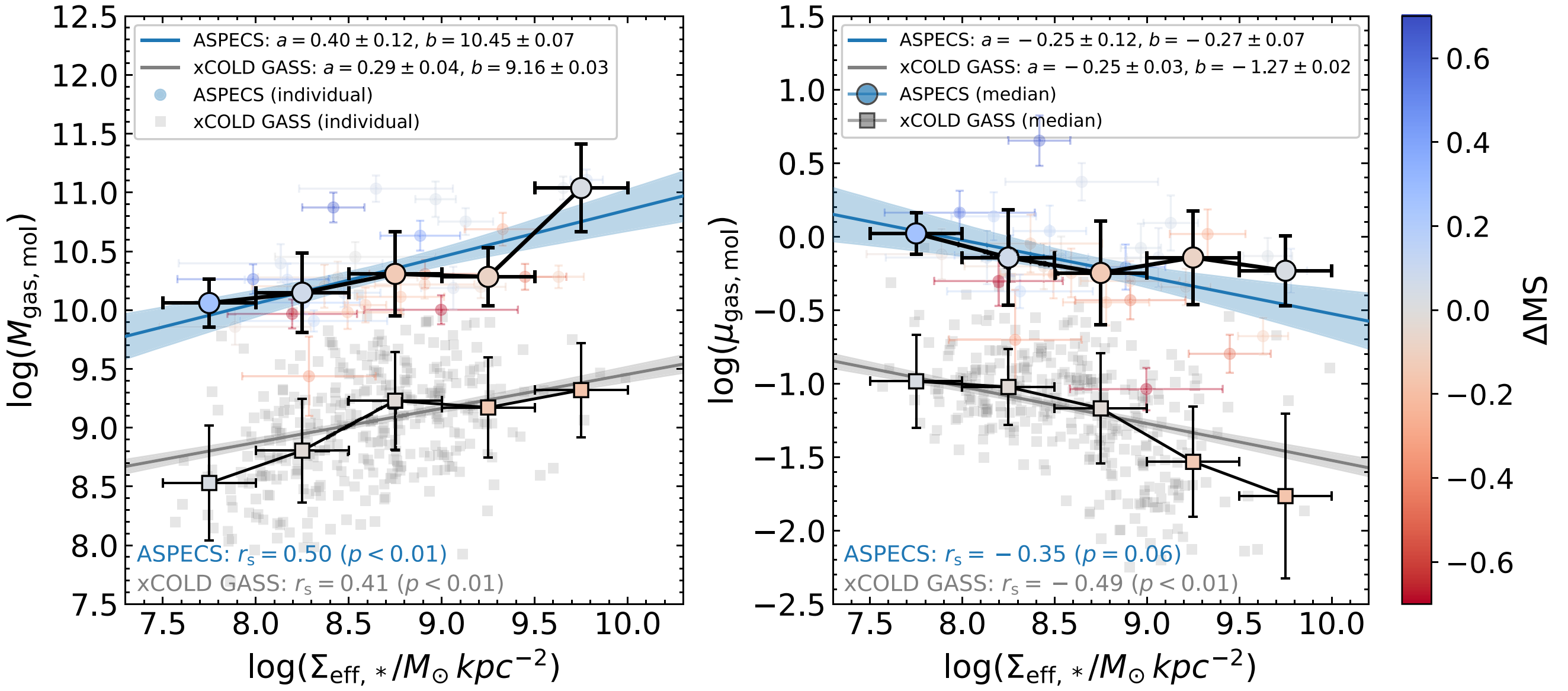

**Figure 12.** Molecular gas ($M_{\rm gas,mol}$, left) and molecular gas-to-stellar mass ratio ($\mu_{\rm gas,mol}$, right) as a function of effective stellar mass surface density ($\Sigma_{\rm eff,*}$) for our $z \sim 1.85$ ASPECS galaxies and the local ($z \leq 0.05$) xCOLD GASS sample (A. Saintonge et al. 2017). ASPECS data points are colour-coded by their offset from the P. Popesso et al. (2023) SFMS ($\Delta$MS), with 1mm.C23 excluded because the galaxy pair is not spatially deblended. xCOLD GASS measurements are shown as grey squares. Equal-width bins of the ASPECS sample are shown as larger circles, colour-coded by the median $\Delta$MS of each bin; the symbols represent the median values, with vertical error bars indicating the standard deviation and horizontal bars indicating the bin width. The same binning analysis is applied to the xCOLD GASS sample. The Spearman correlation coefficient ($r_s$) and associated $p$-value ($p$), calculated from the individual measurements of both samples, are shown in the bottom-left corner of each panel. Best-fitting relations for the individual data points are shown for both samples (ASPECS in blue and xCOLD GASS in grey), with shaded bands indicating the 68 per cent confidence intervals. The fits adopt the form $y = a \times (x - 9) + b$, with the best-fitting parameters and uncertainties reported in the legend. The consistency of the best-fitting slopes between the two redshift samples suggests that the relative role of $\Sigma_{\rm eff,*}$ in regulating the molecular gas reservoir does not evolve significantly across cosmic time.

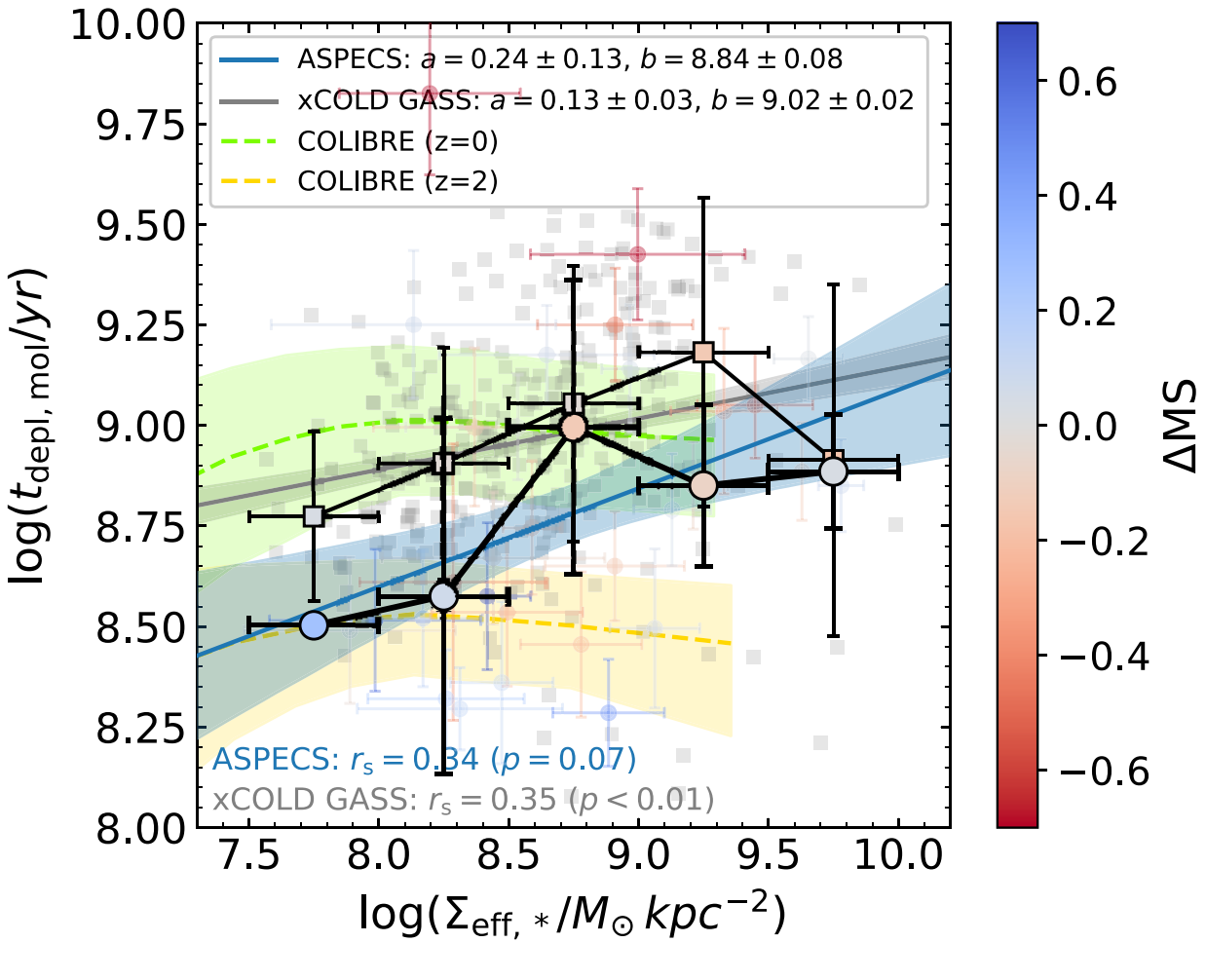

**Figure 13.** Molecular gas depletion time ($t_{\rm depl,mol}$) as a function of effective stellar mass surface density ($\Sigma_{\rm eff,*}$) for our $z \sim 1.85$ ASPECS galaxies and the local ($z \leq 0.05$) xCOLD GASS sample (A. Saintonge et al. 2017). The symbols are the same as those used in Fig. 12. 1mm.C23 is excluded from the plot due to the spatial overlap of the galaxy pair. The green and yellow dashed lines show the $z = 0$ and $z = 2$ predictions from the COLIBRE hydrodynamic simulations (C. d. P. Lagos et al. 2026), respectively, with the shaded regions indicating the 16$^{\rm th}$–84$^{\rm th}$ percentile ranges. The similar trends at $z \sim 0$ and $z \sim 2$ seen in both observations and simulations suggest that the relative role of $\Sigma_{\rm eff,*}$ in regulating $t_{\rm depl}$ does not evolve significantly from cosmic noon to the present day.

We demonstrate that these relations, previously established only for local galaxies (A. K. Leroy et al. 2008; Y. Shi et al. 2011; A. Boselli et al. 2014; M.-L. Huang & G. Kauffmann 2014; A. Saintonge et al. 2017; B. Catinella et al. 2018; J. Freundlich et al. 2019; V. Villanueva et al. 2021; A. Saintonge & B. Catinella 2022; U. Lisenfeld et al. 2023; L. Neumann et al. 2025), are already present at cosmic noon, though the correlations are not as robust as those observed in the nearby Universe due to the smaller sample size. With the advent of *JWST* and high-resolution ALMA observations, future studies of larger high-$z$ samples will be crucial for strengthening these correlations and for advancing our understanding of star formation mechanisms in the early Universe.

## 6 SUMMARY

By combining high-resolution data from *JWST*, *HST*, and ALMA, we perform spatially resolved SED modelling using MAGPHYS to unveil the kpc-scale physical properties of the flux-complete ASPECS galaxies at $0.5 \lesssim z \lesssim 3.7$. We also conduct spatially unresolved SED modelling by incorporating these high-resolution data sets with ancillary observations from *Herschel* and the previous ASPECS-LP ALMA dust continuum measurements. We test multiple SED-modelling configurations (Sections 3.2.2 and 3.2.3) to assess how different modelling choices affect the derived physical properties, providing a practical reference for future studies. Furthermore, we investigate the correlations between global molecular gas properties and $\Sigma_{\rm eff,*}$ for galaxies beyond the local Universe for the first time, offering an alternative perspective on investigating the star formation mechanisms for cosmic noon galaxies. Below, we summarize the main results of this work.

(i) The composite SEDs show systematic trends in dust attenuation and $T_{\rm dust}$ across the redshift bins, which are primarily driven by the $L_{\rm IR}$–$z$ correlation in our sample. When normalized to the same rest-frame 350 μm luminosity, the composite SED of the ASPECS sample is in close agreement with those of local starburst galaxies such as M82, suggesting broadly similar dust attenuation and re-emission properties.

(ii) Across different configurations varying wavelength coverage, spatial resolution, and adopted MAGPHYS models, we find that the derived $M_*$ are robust, with deviations of $\lesssim 0.1$ dex under the MAGPHYS framework. We do not observe a significant 'outshining' effect as reported in previous studies (M. Sawicki & H. K. C. Yee 1998; S. Wuyts et al. 2012; R. Sorba & M. Sawicki 2015, 2018). This likely results from our inclusion of rest-frame NIR data in the spatially resolved modelling, consistent with the finding of J. Song et al. (2023) that stellar mass discrepancies become negligible once rest-frame wavelengths beyond $\sim 1$ μm are covered, for our modest dust-attenuated galaxies.

(iii) In contrast, dust-sensitive properties such as SFR, $L_{\rm dust}$, and $M_{\rm dust}$ are more sensitive to the adopted SED-modelling setups. Compared to the spatially unresolved modelling, which includes additional constraints from *Herschel* and ALMA data, these quantities can be overestimated by up to ~1.0 dex when using the default MAGPHYS models. This result is consistent with U. Dudzevičiūtė et al. (2020) and C. Pacifici et al. (2023) and emphasizes the critical role of MIRI and ALMA data in achieving reliable spatially resolved SED modelling.

(iv) Given the limited availability of high-resolution MIR and FIR data, we introduce an alternative approach that restricts the MAGPHYS model library based on spatially unresolved FIR colours to better constrain the dust SED. By removing less relevant models according to these colours, we find that dust-related quantities are systematically reduced, with the summed dust luminosities agreeing with the spatially unresolved fittings to $\lesssim$ 0.1 dex. This method provides a practical way to obtain more robust dust parameters when spatially resolved FIR constraints are not available.

(v) The MIRI data play a critical role in our spatially resolved modelling, as their inclusion substantially improves the constraints on derived physical properties. This improvement primarily arises from the extended wavelength coverage provided by MIRI, highlighting the importance of rest-frame NIR observations for robust model constraints.

(vi) The FIR constraints from ALMA are equally important for deriving reliable spatially resolved properties. Including ALMA data not only helps reduce the overestimation of the dust-related properties but also provides valuable information in the Rayleigh–Jeans regime, which helps mitigate the age–dust degeneracy. Consistent with the findings of J. Li et al. (2024), we find that the derived $A_{\rm V}$, SFR, and $L_{\rm dust}$ in the galactic centre are systematically lower when ALMA data are included.

(vii) From the curve-of-growth analysis of the $M_*$ maps and *JWST* imaging, we measure the effective radii of these profiles and derive the corresponding $\Sigma_{\rm eff,*}$. We find that the sizes decrease with increasing wavelength, reflecting a combination of intrinsic wavelength-dependent structure and the effects of dust obscuration. In addition, the sizes measured from the rest-frame NIR light trace those derived from the $M_*$ distributions, indicating that the rest-frame NIR light is a reasonable proxy for the stellar mass structure for our ASPECS galaxies.

(viii) By examining the relations between $\Sigma_{\rm eff,*}$ and three molecular gas properties ($M_{\rm gas,mol}$, $\mu_{\rm gas,mol}$, and $t_{\rm depl,mol}$), we find that, at a fixed $\Sigma_{\rm eff,*}$, the ASPECS galaxies exhibit higher $M_{\rm gas,mol}$, higher $\mu_{\rm gas,mol}$, and shorter $t_{\rm depl,mol}$ compared to their local counterparts. In addition, the best-fitting slopes in the local and cosmic noon samples show no significant differences, indicating that we cannot rule out a similar role of $\Sigma_{\rm eff,*}$ in regulating the relative change of gas reservoir, gas-to-stellar mass ratio, and star formation efficiency from cosmic noon to the present. We note, however, that the absolute normalization of these results is sensitive to the choice of $M_{\rm gas,mol}$ conversion factors (e.g. $\kappa_\nu$, $\delta_{\rm GDR}$ and $\alpha_{\rm CO}$, etc.), though the overall trends are likely less affected.

Expanding this analysis to galaxies over a broader redshift range will be important for better constraining the gas–$\Sigma_{\rm eff,*}$ relations presented in Fig. 12 and 13, and to further investigate star formation mechanisms from the perspective of stellar potential. Future facilities such as PRIMA, capable of probing the warm-dust emission at $\lambda_{\rm rest} \sim 20$ μm in cosmic noon galaxies, will also be valuable for better constraining their SEDs in this poorly sampled wavelength regime. Furthermore, given the potential overestimation of dust-sensitive quantities and the persistence of the age–dust degeneracy when FIR constraints are limited, future high-resolution FIR observations will be essential for accurately characterizing the physical conditions of galaxies at high redshift. In particular, the upcoming ALMA Large Program HIgh-definition Dust Imaging of Normal Galaxies in the Hubble Ultra Deep Field (HIDING; PID: 2025.1.01377.L), which will map the rest-frame 370 μm dust continuum of ASPECS galaxies at ~1 kpc ($\sim$ 0.2 arcsec) resolution, will provide impactful insights into the spatially resolved dust and star formation physics of galaxies at cosmic noon. In addition, upgrades to current (sub-)millimetre facilities, together with the development of next-generation observatories, will improve observational efficiency and further advance the field.

## ACKNOWLEDGEMENTS

We thank the anonymous referee for feedback on the manuscript, which helped improve the paper. C-LL, LAB, and JAH acknowledge support from the European Research Council (ERC) Consolidator Grant 101 088 676 ('VOYAJ'). LAB acknowledges support from the Dutch Research Council (NWO) under Veni grant VI.Veni.242.055 (https://doi.org/10.61686/LAJVP77714). PS is supported by the Leiden University Oort Fellowship and the International Astronomical Union–Gruber Foundation Fellowship. MA is supported by FONDECYT grant number 1252054, and gratefully acknowledges support from ANID Basal Project FB210003 and ANID MILENIO NCN2024_112. MK acknowledges support from the Australian Research Council through the Discovery Early Career Researcher Award DE250100709. IS acknowledges funding from STFC (ST/X001075/1). RD is grateful for support from the INAF RSN1 minigrant 2024 'The interstellar medium at high redshift'. DR gratefully acknowledges support from the Collaborative Research Center 1601 (SFB 1601 sub-projects C1, C2, C3, and C6) and through the Cluster of Excellence 'Our Dynamic Universe' under Germany's Excellence Strategy, both funded by the Deutsche Forschungsgemeinschaft (DFG)–500700252, and EXC 3037–533607693. The authors acknowledge assistance from Allegro, the European ALMA Regional Center node in the Netherlands. This paper makes use of the following ALMA data: ADS/JAO.ALMA#2012.1.00173.S, ADS/JAO.ALMA#2016.1.00324.S, and ADS/JAO.ALMA#2021.1.00104.S. ALMA is a partnership of ESO (representing its member states), NSF (USA) and NINS (Japan), together with NRC (Canada), NSTC and ASIAA (Taiwan), and KASI (Republic of Korea), in cooperation with the Republic of Chile. The Joint ALMA Observatory is operated by ESO, AUI/NRAO and NAOJ. This work is based on observations made with the NASA/ESA/CSA *JWST*. The data were obtained from the Mikulski Archive for Space Telescopes at the Space Telescope Science Institute, which is operated by the Association of Universities for Research in Astronomy, Inc., under NASA contract NAS 5-03127. The *JWST* data products presented in this work were retrieved from the Dawn *JWST* Archive (DJA). DJA is an initiative of the DAWN, which is funded by the Danish National Research Foundation under grant DNRF140.

## DATA AVAILABILITY

The *HST* and *JWST* data used for this work can be accessed via the DJA Website: https://dawn-cph.github.io/dja/index.html, DOI: 10.5281/zenodo.15535600, and DOI: 10.17909/et3f-zd57. The ALMA data used for this work are available on the ALMA Science Archive at https://almascience.nrao.edu/aq/ (project codes: 2012.1.00173.S, 2016.1.00324.S, and 2021.1.00104.S).

## SUPPLEMENTARY MATERIAL

Supplementary data are available at *MNRAS* online.

**suppl_data**

Please note: Oxford University Press is not responsible for the content or functionality of any supporting materials supplied by the authors. Any queries (other than missing material) should be directed to the corresponding author for the article.

## APPENDIX A: RESULTS FROM SPATIALLY UNRESOLVED SED MODELLING

The physical properties derived from spatially unresolved SED modelling are summarized in Table A1. A detailed discussion of these results is provided in Section 4.2.1.

**Table A1.** Physical properties derived from the spatially unresolved SED modelling.

| ID (1) | $\log_{10}(M_*/\mathrm{M}_\odot)$ (2) | $\log_{10}(\mathrm{SFR}/\mathrm{M}_\odot\,\mathrm{yr}^{-1})$ (3) | $A_V$ (4) | $\log_{10}(\mathrm{Age/yr})$ (5) | $\log_{10}(L_{\mathrm{dust}}/\mathrm{L}_\odot)$ (6) | $\log_{10}(M_{\mathrm{dust}}/\mathrm{M}_\odot)$ (7) | $T_{\mathrm{dust}}/K$ (8) |
|---|---|---|---|---|---|---|---|
| 1mm.C01 | $10.22 \pm 0.15$ | $2.30 \pm 0.17$ | $2.39 \pm 0.04$ | $7.45 \pm 0.01$ | $12.65 \pm 0.03$ | $8.39 \pm 0.09$ | $53 \pm 7$ |
| 1mm.C02 | $10.66 \pm 0.08$ | $1.86 \pm 0.06$ | $1.19 \pm 0.10$ | $8.71 \pm 0.11$ | $11.87 \pm 0.06$ | $8.53 \pm 0.12$ | $40 \pm 4$ |
| 1mm.C03 | $11.02 \pm 0.09$ | $1.80 \pm 0.07$ | $2.46 \pm 0.12$ | $9.00 \pm 0.15$ | $11.93 \pm 0.05$ | $8.43 \pm 0.17$ | $38 \pm 5$ |
| 1mm.C04 | $10.66 \pm 0.19$ | $1.96 \pm 0.10$ | $2.74 \pm 0.36$ | $8.60 \pm 0.24$ | $12.04 \pm 0.08$ | $8.19 \pm 0.12$ | $39 \pm 6$ |
| 1mm.C05 | $11.17 \pm 0.09$ | $1.87 \pm 0.04$ | $1.76 \pm 0.14$ | $9.17 \pm 0.09$ | $11.98 \pm 0.04$ | $8.52 \pm 0.11$ | $37 \pm 5$ |
| 1mm.C06 | $11.34 \pm 0.15$ | $2.26 \pm 0.09$ | $3.71 \pm 0.32$ | $8.88 \pm 0.13$ | $12.41 \pm 0.05$ | $8.60 \pm 0.08$ | $36 \pm 3$ |
| 1mm.C07 | $10.90 \pm 0.08$ | $1.68 \pm 0.13$ | $1.34 \pm 0.19$ | $8.99 \pm 0.10$ | $11.77 \pm 0.12$ | $7.44 \pm 0.14$ | $47 \pm 6$ |
| 1mm.C08 | $11.28 \pm 0.14$ | $2.55 \pm 0.11$ | $1.89 \pm 0.17$ | $8.65 \pm 0.26$ | $12.60 \pm 0.10$ | $7.71 \pm 0.05$ | $64 \pm 4$ |
| 1mm.C09 | $10.11 \pm 0.16$ | $1.79 \pm 0.21$ | $1.74 \pm 0.42$ | $8.23 \pm 0.40$ | $11.83 \pm 0.25$ | $7.71 \pm 0.17$ | $48 \pm 10$ |
| 1mm.C10 | $10.84 \pm 0.11$ | $2.35 \pm 0.05$ | $1.39 \pm 0.09$ | $8.42 \pm 0.16$ | $12.36 \pm 0.05$ | $7.95 \pm 0.09$ | $49 \pm 4$ |
| 1mm.C11 | $10.67 \pm 0.12$ | $1.65 \pm 0.20$ | $1.76 \pm 0.31$ | $8.84 \pm 0.19$ | $11.75 \pm 0.17$ | $8.19 \pm 0.16$ | $36 \pm 6$ |
| 1mm.C12 | $10.10 \pm 0.10$ | $1.75 \pm 0.16$ | $1.56 \pm 0.12$ | $8.79 \pm 0.16$ | $11.44 \pm 0.05$ | $7.75 \pm 0.15$ | $35 \pm 2$ |
| 1mm.C13 | $10.96 \pm 0.10$ | $1.40 \pm 0.09$ | $2.71 \pm 0.14$ | $9.22 \pm 0.14$ | $11.63 \pm 0.05$ | $7.86 \pm 0.08$ | $37 \pm 4$ |
| 1mm.C14a | $10.33 \pm 0.12$ | $1.69 \pm 0.11$ | $2.11 \pm 0.27$ | $8.57 \pm 0.19$ | $11.80 \pm 0.10$ | $7.55 \pm 0.12$ | $54 \pm 5$ |
| 1mm.C14b | $10.19 \pm 0.12$ | $1.74 \pm 0.09$ | $1.09 \pm 0.16$ | $8.37 \pm 0.20$ | $11.76 \pm 0.10$ | $7.55 \pm 0.10$ | $46 \pm 5$ |
| 1mm.C15 | $11.08 \pm 0.08$ | $1.23 \pm 0.09$ | $1.49 \pm 0.12$ | $9.24 \pm 0.11$ | $11.51 \pm 0.05$ | $7.77 \pm 0.12$ | $37 \pm 4$ |
| 1mm.C16 | $10.60 \pm 0.10$ | $1.46 \pm 0.05$ | $0.59 \pm 0.11$ | $9.00 \pm 0.17$ | $11.44 \pm 0.05$ | $7.94 \pm 0.14$ | $35 \pm 5$ |
| 1mm.C17 | $10.52 \pm 0.08$ | $1.57 \pm 0.08$ | $0.89 \pm 0.12$ | $8.84 \pm 0.17$ | $11.57 \pm 0.08$ | $7.70 \pm 0.14$ | $42 \pm 8$ |
| 1mm.C18 | $10.48 \pm 0.10$ | $1.61 \pm 0.12$ | $0.51 \pm 0.16$ | $8.77 \pm 0.18$ | $11.54 \pm 0.15$ | $7.76 \pm 0.16$ | $43 \pm 9$ |
| 1mm.C19 | $10.56 \pm 0.07$ | $1.66 \pm 0.14$ | $0.99 \pm 0.20$ | $8.79 \pm 0.13$ | $11.67 \pm 0.15$ | $7.52 \pm 0.16$ | $54 \pm 7$ |
| 1mm.C20 | $10.74 \pm 0.07$ | $1.06 \pm 0.09$ | $1.24 \pm 0.12$ | $9.34 \pm 0.12$ | $11.20 \pm 0.07$ | $7.79 \pm 0.14$ | $35 \pm 5$ |
| 1mm.C21 | $10.24 \pm 0.10$ | $1.45 \pm 0.15$ | $0.86 \pm 0.20$ | $8.70 \pm 0.20$ | $11.43 \pm 0.16$ | $7.37 \pm 0.15$ | $46 \pm 9$ |
| 1mm.C22 | $10.28 \pm 0.08$ | $1.61 \pm 0.06$ | $1.26 \pm 0.10$ | $8.60 \pm 0.13$ | $11.63 \pm 0.05$ | $7.40 \pm 0.10$ | $49 \pm 5$ |
| 1mm.C23 | $10.68 \pm 0.11$ | $1.49 \pm 0.13$ | $0.41 \pm 0.16$ | $9.02 \pm 0.19$ | $11.40 \pm 0.18$ | $7.48 \pm 0.53$ | $45 \pm 11$ |
| 1mm.C24 | $10.12 \pm 0.12$ | $1.74 \pm 0.13$ | $1.19 \pm 0.22$ | $8.29 \pm 0.27$ | $11.73 \pm 0.15$ | $7.74 \pm 0.16$ | $46 \pm 9$ |
| 1mm.C25 | $10.54 \pm 0.10$ | $1.35 \pm 0.06$ | $1.39 \pm 0.11$ | $9.05 \pm 0.16$ | $11.43 \pm 0.05$ | $7.68 \pm 0.11$ | $37 \pm 4$ |
| 1mm.C26 | $10.29 \pm 0.10$ | $1.30 \pm 0.10$ | $1.41 \pm 0.17$ | $8.86 \pm 0.16$ | $11.37 \pm 0.10$ | $7.62 \pm 0.13$ | $40 \pm 6$ |
| 1mm.C28 | $10.54 \pm 0.11$ | $1.15 \pm 0.12$ | $0.51 \pm 0.17$ | $9.16 \pm 0.20$ | $11.16 \pm 0.13$ | $7.88 \pm 0.20$ | $32 \pm 4$ |
| 1mm.C30 | $9.69 \pm 0.15$ | $1.10 \pm 0.09$ | $1.04 \pm 0.16$ | $8.52 \pm 0.23$ | $11.09 \pm 0.08$ | $6.92 \pm 0.18$ | $53 \pm 8$ |
| 1mm.C31 | $9.98 \pm 0.10$ | $1.37 \pm 0.13$ | $0.31 \pm 0.16$ | $8.48 \pm 0.20$ | $11.22 \pm 0.18$ | $7.35 \pm 0.18$ | $44 \pm 10$ |
| 1mm.C32 | $10.27 \pm 0.15$ | $0.14 \pm 0.21$ | $1.61 \pm 0.21$ | $9.36 \pm 0.22$ | $10.54 \pm 0.06$ | $7.43 \pm 0.14$ | $31 \pm 5$ |
| 1mm.C33 | $11.04 \pm 0.10$ | $0.58 \pm 0.14$ | $0.31 \pm 0.07$ | $9.41 \pm 0.12$ | $10.74 \pm 0.07$ | $7.49 \pm 0.14$ | $33 \pm 5$ |
| 3mm.09 | $11.66 \pm 0.12$ | $2.45 \pm 0.07$ | $4.34 \pm 0.29$ | $8.96 \pm 0.11$ | $12.61 \pm 0.03$ | $8.49 \pm 0.04$ | $40 \pm 2$ |
| 3mm.11 | $10.14 \pm 0.08$ | $0.83 \pm 0.12$ | $0.51 \pm 0.16$ | $9.11 \pm 0.17$ | $10.80 \pm 0.14$ | $6.93 \pm 0.51$ | $44 \pm 11$ |
| 3mm.16 | $10.25 \pm 0.10$ | $1.21 \pm 0.10$ | $0.59 \pm 0.16$ | $8.93 \pm 0.17$ | $11.17 \pm 0.12$ | $7.69 \pm 0.23$ | $37 \pm 8$ |

*Note.* (1) ALMA 1 mm source ID (1mm.C∗: M. Aravena et al. 2020; J. González-López et al. 2020; 3mm.∗: L. A. Boogaard et al. 2019; J. González-López et al. 2019); (2) stellar mass; (3) SFR; (4) $V$-band attenuation; (5) mass-weighted stellar age; (6) dust luminosity; (7) dust mass; (8) dust temperature. A machine-readable version of this table is available in the online supplementary material.

## APPENDIX B: UNCERTAINTY MAPS OF PHYSICAL PROPERTIES

Fig. B1 presents the physical property maps and their corresponding uncertainty maps derived from the restricted-model F444W_hn SED modelling. Equivalent maps from the F1280W_hnm and F1280W_hnma modelling for the full sample are available in the online supplementary material.

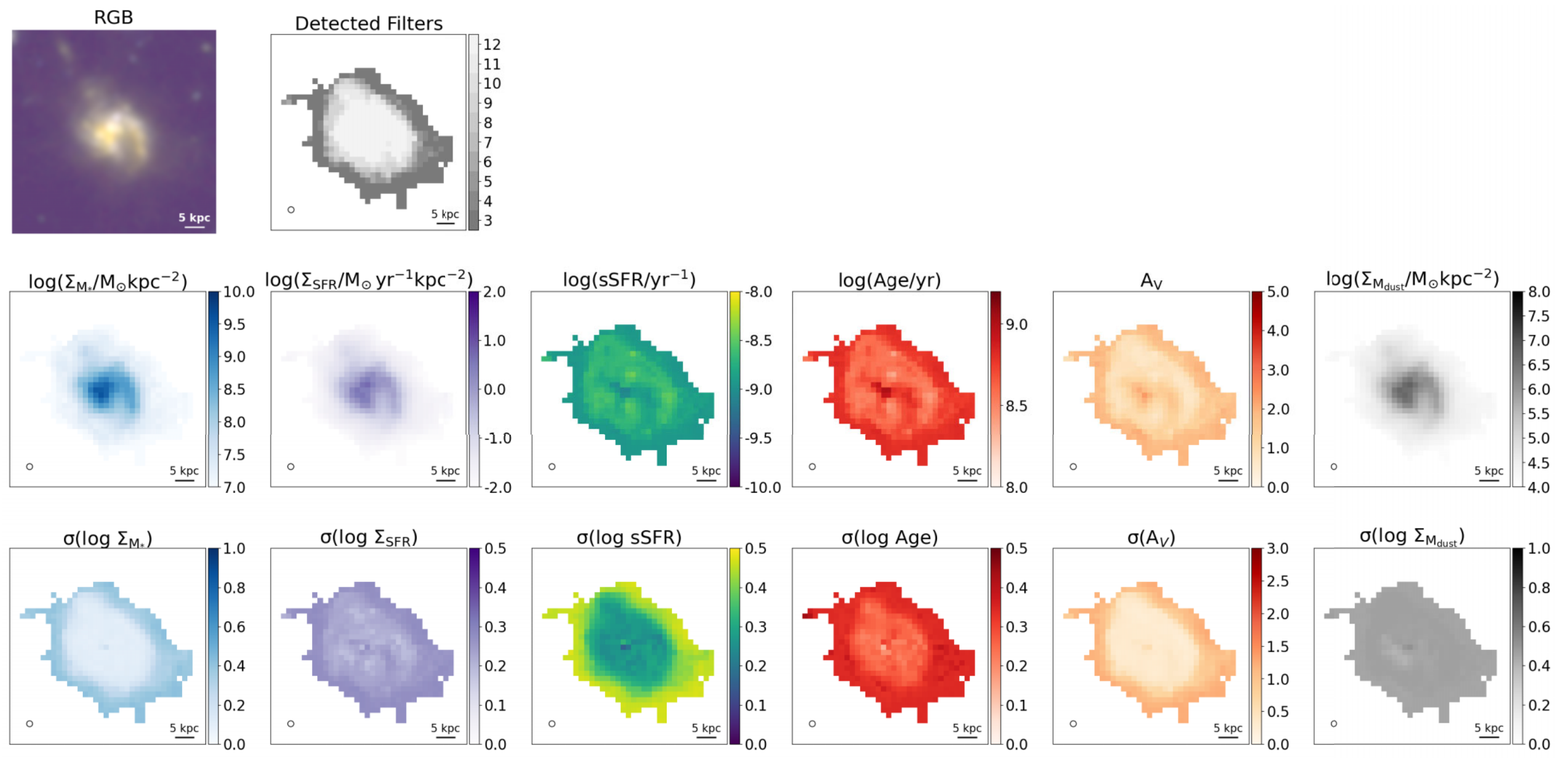

**Figure B1.** Physical property maps and their corresponding uncertainty maps for the example galaxy 1mm.C10, derived from the restricted-model F444W_hn SED modelling and corresponding to the green curves shown in Fig. 5. The first two rows present the same physical property maps as in Fig. 8, the third row shows the corresponding uncertainty maps. The physical properties are less well constrained in the galaxy outskirts due to the lower S/N. Equivalent maps from all modelling configurations (F444W_hn, F1280W_hnm, and F1280W_hnma) for the full sample are available online.

## APPENDIX C: COMPARISON OF DIFFERENT MOLECULAR GAS MASS ESTIMATES

In our sample, 16 out of the 35 galaxies have single or multiple emission-line detections (M. Aravena et al. 2019; L. A. Boogaard et al. 2019; 2020; J. González-López et al. 2019), enabling independent estimates of the molecular gas mass ($M_{\rm mol}$) based on different tracers. In this section, we present the molecular gas masses derived from dust-based SED modelling ($M_{\rm mol,dust}$), the CO-based molecular gas masses reported in the literature ($M_{\rm mol,CO}$), and a comparison between the two methods.

The dust-based molecular gas mass is estimated as

$$M_{\rm mol,dust} = \delta_{\rm GDR} \times M_{\rm dust}, \quad (C1)$$

where $\delta_{\rm GDR}$ represents the molecular gas-to-dust ratio. We adopt the dust masses derived from the spatially unresolved SED fitting, as the inclusion of FIR photometry provides more robust constraints on $M_{\rm dust}$. Instead of the dust mass absorption coefficient adopted in MAGPHYS ($\kappa_{850\,\mu\rm m} = 0.77\,{\rm g}^{-1}\,{\rm cm}^2$; L. Dunne et al. 2000), we use a lower value of $\kappa_{850\,\mu\rm m} = 0.4699\,{\rm g}^{-1}\,{\rm cm}^2$ (B. T. Draine & A. Li 2007; B. T. Draine et al. 2014). This choice results in systematically higher dust masses for a given luminosity and yields better agreement, on average, with the CO-based molecular gas mass estimates. We assume a fixed gas-to-dust ratio of $\delta_{\rm GDR} = 200$, following M. Aravena et al. (2020).

For the CO-based molecular gas mass, we adopt the values provided on the ASPECS website[6], which are described in detail in L. A. Boogaard et al. (2020). These estimates are computed as

$$M_{\rm mol,CO} = \alpha_{\rm CO} \times \frac{L'_{\rm CO(J\to J-1)}}{r_{\rm J1}}, \quad (C2)$$

where $\alpha_{\rm CO}$ is the CO-to-molecular-gas conversion factor (including helium), $L'_{\rm CO(J\to J-1)}$ is the CO luminosity of the observed $J \to J-1$ transition, and $r_{\rm J1}$ represents the line ratio relative to CO(1–0). The adopted conversion factor is $\alpha_{\rm CO} = 3.6\,{\rm M}_\odot\,({\rm K\,km\,s^{-1}\,pc^2})^{-1}$, and redshift-dependent $r_{\rm J1}$ values are applied. Further details can be found in L. A. Boogaard et al. (2020).

Fig. C1 compares the molecular gas masses derived using the two methods for the galaxies with CO measurements. A clear correlation is observed between $M_{\rm mol,dust}$ and $M_{\rm mol,CO}$, indicating overall consistency between the molecular gas mass estimates obtained from different tracers. We do find a median offset of $-0.17 \pm 0.08$ dex between the two methods, which might be attributed to systematic uncertainties associated with the assumed $\kappa_\nu$, $\delta_{\rm GDR}$, $\alpha_{\rm CO}$, and, to a lesser extent, $r_{\rm J1}$. In particular, we note that independent kinematic constraints on the $M_{\rm mol}$ for some sources may be in tension with our choices of $\kappa_\nu$ and $\delta_{\rm GDR}$. An in-depth analysis is beyond the scope of this paper, but overall they do not impact the conclusions, and we note a similar trend was reported in M. Aravena et al. (2020, Appendix A), where a more detailed discussion of molecular gas mass estimates is presented.

[6] https://aspecs.info/

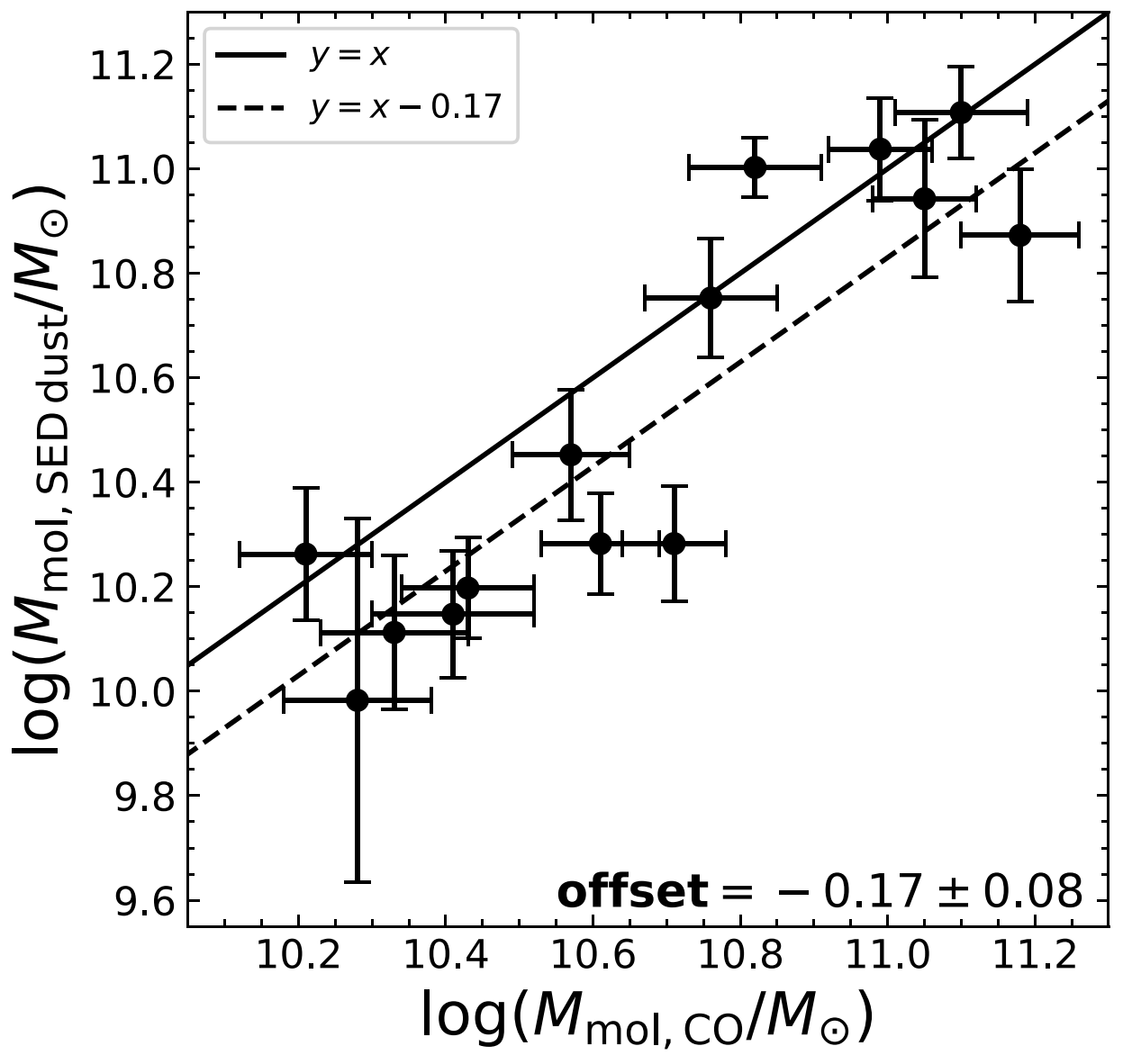

**Figure C1.** Comparison of molecular gas mass obtained using different methods for the galaxies with emission-line detections. The CO-based and SED dust-based molecular gas masses are shown on the abscissa and ordinate, respectively. The solid line indicates the one-to-one relation, while the dashed line represents a 0.17 dex offset below it. The median values and bootstrap uncertainties of the difference between the two estimates are indicated in the bottom-right corner. Overall, the molecular gas masses derived from the two methods are broadly consistent within the uncertainties.

[1]*Leiden Observatory, Leiden University, PO Box 9513, NL-2300 RA Leiden, the Netherlands*
[2]*International Centre for Radio Astronomy Research (ICRAR), The University of Western Australia, M468, 35 Stirling Highway, Crawley WA 6009, Australia*
[3]*ARC Centre of Excellence for All Sky Astrophysics in 3 Dimensions (ASTRO 3D), Australia*
[4]*International Space Centre (ISC), The University of Western Australia, M468, 35 Stirling Highway, Crawley WA 6009, Australia*
[5]*Research School of Astronomy and Astrophysics, Australian National University, Cotter Road, Weston Creek ACT 2611, Australia*
[6]*Centre for Extragalactic Astronomy, Department of Physics, Durham University, South Road, Durham DH1 3LE, UK*
[7]*Instituto de Estudios Astrofísicos, Facultad de Ingeniería y Ciencias, Universidad Diego Portales, Av. Ejército 441, Santiago, Chile*
[8]*Millennium Nucleus for Galaxies (MINGAL), Santiago, Chile*
[9]*Instituto de Estudios Astrofísicos, Facultad de Ingeniería y Ciencias, Universidad Diego Portales, Av. Ejército 441, Santiago 8370191, Chile*
[10]*Jodrell Bank Centre for Astrophysics, University of Manchester, Oxford Road, Manchester M13 9PL, UK*
[11]*Centro de Astrobiología (CAB), CSIC-INTA, Ctra. de Ajalvir km 4, Torrejón de Ardoz, E-28850 Madrid, Spain*
[12]*Institut d'Astrophysique de Paris, Sorbonne Université, UPMC Université Paris 6 and CNRS, UMR 7095, 98 bis boulevard Arago, F-75014 Paris, France*
[13]*INAF–Osservatorio di Astrofisica e Scienza dello Spazio di Bologna, Via Gobetti 93/3, I-40129 Bologna, Italy*
[14]*Institute of Astrophysics, Foundation for Research and Technology–Hellas (FORTH), Heraklion GR-70013, Greece*
[15]*European Southern Observatory, Karl-Schwarzschild-Str. 2, D-85748 Garching, Germany*
[16]*Hiroshima Astrophysical Science Center, Hiroshima University, 1-3-1 Kagamiyama, Higashi-Hiroshima, Hiroshima 739-8526, Japan*
[17]*Université Paris-Saclay, Université Paris Cité, CEA, CNRS, AIM, F-91191 Gif-sur-Yvette, France*
[18]*Department of Astronomy, Oscar Klein Centre, Stockholm University, AlbaNova, SE-106 91 Stockholm, Sweden*
[19]*National Radio Astronomy Observatory, 520 Edgemont Rd, Charlottesville, VA 22903, USA*
[20]*Institut für Astrophysik, Universität zu Köln, Zülpicher Straße 77, D-50937 Köln, Germany*
[21]*Max-Planck-Institut für Astronomie, Königstuhl 17, D-69117 Heidelberg, Germany*
[22]*Max-Planck-Institut für Radioastronomie, Auf dem Hügel 69, D-53121 Bonn, Germany*